\documentclass[amsmath,amssymb,aps,reprint]{revtex4-1} % chktex 8

\usepackage{iftex}

\ifPDFTeX%
\usepackage[T1]{fontenc}
\usepackage[utf8]{inputenc}

\usepackage[hidelinks,colorlinks,breaklinks]{hyperref}

\usepackage{cleveref}

\newcommand{\vect}[1]{\mathbf{#1}}

\crefname{equation}{equation}{equations}
\crefname{chapter}{chapter}{chapters}
\crefname{section}{section}{sections}
\crefname{appendix}{appendix}{appendices}
\crefname{enumi}{item}{items}
\crefname{footnote}{footnote}{footnotes}
\crefname{figure}{figure}{figures}
\crefname{table}{table}{tables}
\crefname{theorem}{theorem}{theorems}
\crefname{lemma}{lemma}{lemmas}
\crefname{corollary}{corollary}{corollaries}
\crefname{proposition}{proposition}{propositions}
\crefname{definition}{definition}{definitions}
\crefname{result}{result}{results}
\crefname{example}{example}{examples}
\crefname{remark}{remark}{remarks}
\crefname{note}{note}{notes}
 \else\ifXeTeX%
\providecommand{\symbf}{\mathbf}

\usepackage{fontspec}

\usepackage{unicode-math}

\usepackage{microtype}

\usepackage[hidelinks,colorlinks]{hyperref}

\usepackage[noabbrev]{cleveref}

\newcommand{\vect}[1]{\symbf{#1}}
 \fi\fi%

\usepackage{amsmath}
\usepackage{amssymb}
\usepackage{mathtools}
\usepackage{siunitx}
\usepackage{graphicx}
\usepackage[english]{babel}
\usepackage{csquotes}

\usepackage{titlesec}

\usepackage{cool}

\newcommand{\creflastconjunction}{, and\nobreakspace}

\usepackage[version=3]{mhchem}

\usepackage{braket}

\newcounter{intropara}
\newcommand\introparanum{\par\refstepcounter{intropara}{(\theintropara)}~}

\DeclarePairedDelimiter\abs{\lvert}{\rvert}
\DeclarePairedDelimiter\norm{\lVert}{\rVert}

\makeatletter
\let\oldabs\abs{}
\def\abs{\@ifstar{\oldabs}{\oldabs*}}
\makeatother

\newcommand{\re}{\operatorname{Re}}

\newcommand*\cc[1]{#1^*}

\newcommand{\hc}{\text{h.c.}}

\newcommand{\vc}{\vect}

\newcommand{\vR}{\vect{R}}
\newcommand{\vK}{\vect{k}}

\newcommand\normalorder[1]{{:}\mkern1mu#1\mkern1.6mu{:}}

\newcommand{\s}{s}

\newcommand{\of}[1]{\left(#1\right)}
\newcommand{\ofK}{\of{\vK}}
\newcommand{\ofMK}{\of{-\vK}}
\newcommand{\ev}[1]{\left\langle#1\right\rangle}

\newcommand{\fnEnergy}[1]{E_{{\tau} {\s}}^{#1}}
\newcommand{\fnTheta}[1]{{\theta}_{{\tau} {\s}}^{#1} \of{k}}

\newcommand{\sumK}{{\sum}_{\vK}}
\newcommand{\sumKK}{{\sum}_{\vK, \vK'}}

\newcommand{\ketOrb}[2]{\Ket{v_{{\tau} {\s}}^{#1} #2}}

\newcommand*\dif{\mathop{}\!\mathrm{d}}
 
\begin{document}
  \author{Evan Sosenko}
  \email{evan.sosenko@email.ucr.edu}
  \homepage{https://evansosenko.com}

  \author{Junhua Zhang}
  \email{junhua.zhang@ucr.edu}

  \author{Vivek Aji}
  \email{vivek.aji@ucr.edu}

  \affiliation{%
    Department of Physics, University of California,
    Riverside, Riverside, California 92521, USA}

  \title{%
    Unconventional superconductivity and anomalous response
    in hole-doped transition metal dichalcogenides}
  \date{\today}

  \begin{abstract}
  Two dimensional transition metal dichalcogenides entwine
  interaction, spin-orbit coupling, and topology.
  Hole-doped systems lack spin degeneracy:
  states are indexed with spin and valley specificity.
  This unique structure offers new possibilities
  for correlated phases and phenomena.
  We realize an unconventional superconducting pairing phase
  which is an equal mixture of a spin singlet and the $m = 0$ spin triplet.
  It is stable against large in-plane magnetic fields,
  and its topology allows quasiparticle excitations
  of net nonzero Berry curvature via pair-breaking circularly polarized light.
\end{abstract}
   \maketitle
  \section{Introduction}

The interplay of spin-orbit interaction and electron-electron interaction
is a fertile area of research where new phases of matter
and novel phenomena have been theoretically conjectured
and experimentally realized
\cite{%
  PhysRevLett.61.2015,%
  PhysRevLett.95.226801,%
  PhysRevLett.96.106802,%
  Konig02112007,%
  RevModPhys.82.3045,%
  RevModPhys.83.1057,%
  doi:10.1146/annurev-conmatphys-020911-125138% chktex 8
}.
Single-layer transition metal group-VI dichalcogenides (TMDs),
\ce{MX2} ($\ce{M} = \ce{Mo}, \ce{W}$
and $\ce{X} = \ce{S}, \ce{Se}, \ce{Te}$),
are direct band gap semiconductors that have all the necessary ingredients
to explore these phenomena
\cite{%
  RadisavljevicB.2011,%
  PhysRevB.84.153402,%
  doi:10.1021/nl2021575,%
  Wang2012,%
  Ye30112012,%
  Bao2013,%
  1.4804936,%
  PhysRevB.88.075409,%
  Xu2014,%
  PhysRevB.91.094510,%
  1508.03068%
}.
While sharing the hexagonal crystal structure of graphene,
they differ in three important aspects:
(1) gapped valleys
as opposed to Dirac nodes;
(2) broken inversion symmetry and strong spin-orbit coupling yielding
a large splitting of the valence bands;
and (3) the bands near the chemical potential predominantly have
the transition metal $d$-orbital character
\cite{%
  0022-3719-5-7-007,% chktex 8
  PhysRevB.64.235305,%
  PhysRevLett.105.136805,%
  doi:10.1021/nl903868w,%
  PhysRevB.88.045416,%
  PhysRevB.88.085433%
}.

The inversion symmetry breaking
and the strong spin-orbit coupling due to the heavy transition element
(\ce{Mo} and \ce{W})
endow the bands with nontrivial Berry curvature.
A remarkable consequence is that
spin-preserving optical transitions between valence
and conduction bands are allowed,
even though the atomic orbitals involved all have a $d$-character.
Furthermore, the valley-dependent sign of
the Berry curvature leads to selective photoexcitation:
right circular polarization couples to one valley,
and left circular polarization to the other.
This enables a number of valleytronic and spintronic applications
that have attracted a lot of attention over the last few years
\cite{%
  RevModPhys.82.1959,%
  PhysRevLett.108.196802,%
  Mak27062014%
}.

We are primarily interested in exploiting
the band structure and valley-contrasting probe afforded by
the nontrivial topology in order to study and manipulate
correlated phenomena in these systems.
In particular, we focus on hole-doped systems,
where an experimentally accessible window in energy
is characterized by two disconnected pieces of
spin non-degenerate Fermi surfaces.
One can preferentially excite electrons from either Fermi surface.
Since the spins are locked to their valley index,
these excitations have specific $s_z$
(where the $z$-axis is perpendicular to the two-dimensional crystal).
We focus on the possible superconducting states and their properties.

Spin-valley locking and its consequence for superconductivity,
dubbed Ising superconductivity, has been previously studied
for heavily doped $p$-type and $n$-type TMDs
\cite{%
  Lu1353,%
  Xi2016,%
  Saito2016,%
  PhysRevB.93.180501,%
  PhysRevLett.113.097001%
},
where Fermi surfaces of each spin are present in each valley.
Our focus is the regime of maximal loss of spin degeneracy where the
effects are most striking
\cite{PhysRevB.94.060501}.
The two valleys in the energy landscape generically allow
two classes of superconducting phases:
intervalley pairing with zero center of mass momentum,
and intravalley pairing with finite Cooper pair center of mass.
Since center-of-symmetry is broken and spin degeneracy is lost,
classifications of superconducting states by parity,
i.e., singlet vs.\ triplet, is no longer possible.
In this paper, we study both extrinsic and intrinsic superconductivity
by projecting the interactions and pairing potential to
the topmost valence band.
We identify the possible phases, and analyze the nature
of the optoelectronic coupling and the response to magnetic fields.
Our main conclusions are as follows:

\introparanum{}
For both proximity to an $s$-wave superconductor,
and due to local attractive density-density interactions,
the leading instability is due to an intervalley paired state,
where the Cooper pair is an equal mixture
of a spin singlet and the $m = 0$ spin triplet
\cite{PhysRevLett.87.037004}.

\introparanum{}
While the valley selectivity of the optical transition is suppressed,
it remains finite.
Consequently, the two quasiparticles
generated by pair-breaking circularly polarized light
are correlated such that one is in the valence band of one valley
and the conduction band of the other.
The valley and bands are determined by the polarity of incident light.

\introparanum{}
The quasiparticles generated in (2)
both have the same charge and Berry curvature.
Thus an anomalous Hall effect is anticipated
as the two travel in the same direction transverse to an applied electric field.

\introparanum{}
An in-plane magnetic field tilts the spin,
modifying the internal structure of the Cooper pair,
however, no pair-breaking is induced in the absence of scalar impurities.
The suppression of the effective interaction leads
to a parametric reduction of the transition temperature.
In the presence of scalar impurities, pair-breaking is enabled,
but the associated critical magnetic field is large.
   \section{Model}

The TMD system is described by
the effective tight-biding, low-energy, two-valley Hamiltonian
\cite{PhysRevLett.108.196802},
\begin{equation}
  \label{eq:hamiltonian}
  H_{\tau}^0 \ofK
  = a t \left( {\tau} k_x \hat{{\sigma}}_x + k_y \hat{{\sigma}}_y \right)
  + \frac{E_g}{2} \hat{{\sigma}}_z - E_{\text{soc}} {\tau} \frac{\hat{{\sigma}}_z - 1}{2} \hat{s}_z.
\end{equation}
where the Pauli matrices $\hat{s}_i$ operate in the spin space and
$\hat{{\sigma}}_i$ operate in the orbital space
with the two Bloch orbital states $\ketOrb{{\nu}}{\ofK}$
(indexed by ${\nu} = +$ for the in-plane orbital state
$\Ket{d_{x^2 - y^2}} + i {\tau} \Ket{d_{xy}}$
and ${\nu} = -$ for the out-of-plane orbital state $\Ket{d_{z^2}}$),
${\s} = {\pm}$ is the spin index, and ${\tau} = {\pm}$ is the valley index corresponding
to the ${\pm} \vc{K}$ point, respectively.
The momentum $\vK = \left( k_x, k_y \right)$
is measured from the valley center, $a$ is the lattice constant,
$t$ is the hopping parameter, $E_g$ represents the energy
gap between the conduction and valence bands, and $2E_{\text{soc}}$ is the
spin splitting energy in the valence bands due to spin-orbit interaction.

\begin{figure}
  \includegraphics[width=\columnwidth]{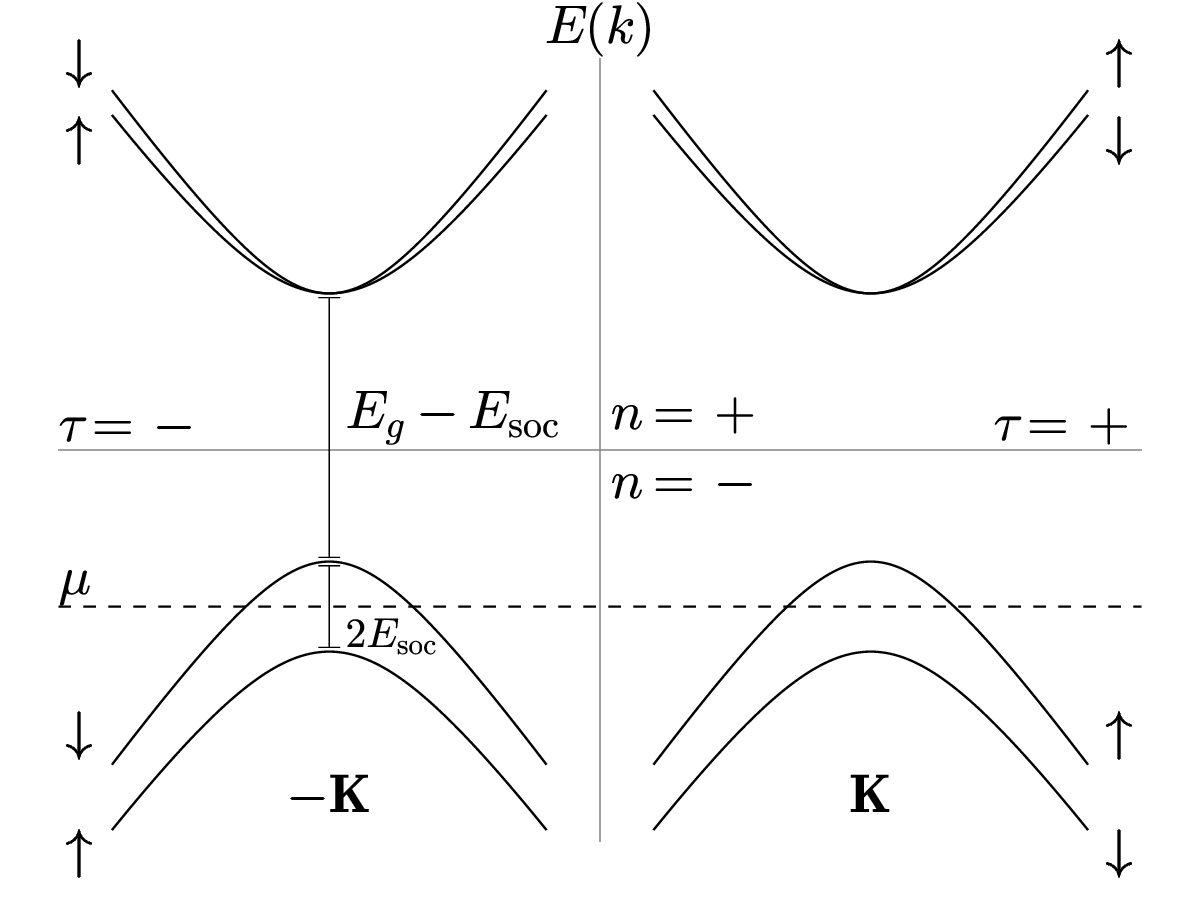}
  \caption{%
    Energy bands for $\ce{WSe2}$ as given by \cref{eq:energy}
    with $a t = \SI{3.939}{\electronvolt \per \angstrom}$,
    $E_g = \SI{1.60}{\electronvolt}$,
    and $E_{\text{soc}} = \SI{0.23}{\electronvolt}$.
    Each valley is centered at ${\pm} \vc{K}$ relative to the center of the
    Brillouin zone.
    The energy for a given band depends only on the distance $k$
    measured from the valley center.
  }\label{fig:energy}
\end{figure}

The energy spectrum,
\begin{equation}
  \label{eq:energy}
  2 \fnEnergy{n} \of{k}
  = {\tau} {\s} E_{\text{soc}} + n \sqrt{{\left( 2 a t k \right)}^2
  + {\left( E_g - {\tau} {\s} E_{\text{soc}} \right)}^2}.
\end{equation}
with $k = \abs{\vK}$
and $n = 1$ ($n = -1$) indexing the conduction (valence) band
is shown in \cref{fig:energy}.

We focus on doped systems
such that the chemical potential ${\mu}$ lies in the upper valence bands.
Within each band, the Bloch basis eigenstates are written
in terms of the orbital states as elements on the Block sphere,
\begin{equation}
  \begin{aligned}
    \Ket{u_{{\tau} {\s}}^n \of{k, {\phi}}}
    = & \cos{\frac{\fnTheta{n}}{2}} \ketOrb{+}{\of{k, {\phi}}} \\
    + e^{-i {\tau} {\phi}}
      & \sin{\frac{\fnTheta{n}}{2}} \ketOrb{-}{\of{k, {\phi}}},
  \end{aligned}
\end{equation}
where $k_x + i {\tau} k_y = k e^{i {\tau} {\phi}}$ and
\begin{equation}
  \tan{\frac{\fnTheta{n}}{2}}
  = \frac{a t {\tau} k}{\dfrac{E_g}{2} - \fnEnergy{-n} \of{k}}
  = \frac{a t {\tau} k}{\fnEnergy{n} \of{k} - \fnEnergy{-} \of{0}}.
\end{equation}
The polar angle on the Bloch sphere
of the conduction and valence bands are related by
$\fnTheta{-} - \fnTheta{+} = {\tau} {\pi}$.
The mapping of the energy band to the Bloch sphere,
parametrized by $\left( {\theta}, {\phi} \right)$,
encodes the topological character:
as one moves from the node out to infinity,
the states sweep either the northern or southern hemisphere
with a chirality determined by the Berry curvature.
   \section{Superconductivity}
\label{s:superconductivity}

We consider two approaches to realizing a superconducting state.
First, we assume a proximity induced state obtained by
layering a TMD on an $s$-wave superconductor.
Second, we study an intrinsic correlated phase arising
from density-density interactions.

We use $d^{\nu}_{{\tau} {\s}} \ofK$ as the annihilation operator
for tight-binding $d$-orbital states,
and $c^n_{{\tau} {\s}} \ofK$ for the eigenstates of the non-interacting Hamiltonian,
${\lambda}_{\vK}$ for the energy dispersion for Bogoliubov quasiparticles,
and ${\Delta}_{\vK}$ for the superconducting gap function.

\subsection{Induced State}

A proximity $s$-wave superconductor will inject Cooper pairs
according to
\begin{equation}
  H^V
  = {\sum}_{\vK, {\nu}, {\tau}} \cc{B}_{\nu}
    d^{\nu}_{-{\tau} {\downarrow}} \ofMK d^{\nu}_{{\tau} {\uparrow}} \ofK + \frac{{\varepsilon}}{2} + \hc
\end{equation}
The coupling constants $B_{\nu}$ and the overall constant ${\varepsilon}$
depend on the material interface %
\footnote{%
  Note that all sums over $\vK$ are restricted to $\left| \vK \right|$
  less than some cutoff that restricts the momentum to a single valley.
}.
Using the abbreviated notation
$c_{\vK {\alpha}} = c^-_{{\tau} {\s}} \ofK$,
with ${\alpha} =\ {\uparrow}{\downarrow}$ for ${\tau} = {\s} = {\pm}$,
projecting onto the upper valence bands yields,
\begin{multline}
  \label{eq:induced}
  P_{{\tau} = {\s}}^{n = -} \left( H^0 + H^V - {\mu} N \right)
  = {\sum}_{\vK, {\alpha}} {\xi}_{\vK} c_{\vK {\alpha}}^{\dagger} c_{\vK {\alpha}} \\
  - \sumK \left( \cc{{\Delta}}_{\vK} c_{-\vK {\downarrow}} c_{\vK {\uparrow}}
  + {\Delta}_{\vK} c_{\vK {\uparrow}}^{\dagger} c_{-\vK {\downarrow}}^{\dagger} \right)
  + {\varepsilon},
\end{multline}
where ${\xi}_{\vK} = E_{+ {\uparrow}}^- \of{\abs{\vK}} - {\mu}$ and
the effective BCS gap function is
\begin{equation}
  {\Delta}_{\vK}
  = \frac{1}{2} \left( B_+ + B_- \right)
  + \frac{1}{2} \left( B_+ - B_- \right)
    \cos{{\theta}_{\vK}},
\end{equation}
with ${\theta}_{\vK} = {\theta}_{+{\uparrow}}^- \of{\abs{\vK}}$.
This form is identical to the standard BCS Hamiltonian with
an effective spin index ${\alpha}$.
However, the spin state of the Cooper pair is an equal superposition
of the singlet and the $m = 0$ component of spin triplet.
The corresponding quasiparticle eigenstates are
${\gamma}_{\vK {\alpha}}
= {\alpha} \cos{{\beta}_{\vK}} c_{\vK {\alpha}} + \sin{{\beta}_{\vK}} c_{-\vK, -{\alpha}}^{\dagger}$,
with energies
${\lambda}_{\vK} = {\pm} \sqrt{{\xi}_{\vK}^2 + {\Delta}_{\vK}^2}$,
where $\cos{2 {\beta}_{\vK}} = {\xi}_{\vK} / {\lambda}_{\vK}$.
Note that if $B_+ = B_-$,
then ${\Delta}_{\vK}$ is a constant and independent of $\vK$.
Even when $B_+$ and $B_-$ are different,
the constant term dominates.
Before exploring the nature of this state,
we analyze the case of intrinsic superconductivity,
and show that the same state is energetically preferred.

\subsection{Intrinsic Phase}

For a local attractive density-density interaction
(e.g.\ one mediated by phonons), the potential is
$V {\simeq} \frac{1}{2} {\sum}_{\vR, \vR'} v_{\vR \vR'}
\normalorder{n_{\vR} n_{\vR'}}$,
with $v_{\vR \vR'} = v_0 {\delta}_{\vR \vR'}$
and $n_{\vR}$ the total Wannier electron density at lattice vector $\vR$.
Projecting onto states near the chemical potential gives
\begin{multline}
  \label{eq:channels}
  P_{{\tau} = {\s}}^{n = -} \left( H^V \right)
  = \sumKK v \of{\vK' - \vK} \\
  {\times} \left(
    A_{\vK \vK'}^2 c_{\vK' {\uparrow}}^{\dagger} c_{-\vK' {\uparrow}}^{\dagger} c_{-\vK {\uparrow}} c_{\vK {\uparrow}}
  + A_{\vK' \vK}^2 c_{\vK' {\downarrow}}^{\dagger} c_{-\vK' {\downarrow}}^{\dagger} c_{-\vK {\downarrow}} c_{\vK {\downarrow}}
    \right. \\ + \left.
      2 \abs{A_{\vK \vK'}}^2
      c_{\vK' {\uparrow}}^{\dagger} c_{-\vK' {\downarrow}}^{\dagger} c_{-\vK {\downarrow}} c_{\vK {\uparrow}}
    \vphantom{2 \abs{V_{\vK \vK'}}^2} \right),
\end{multline}
where
\begin{equation}
  A_{\vK \vK'}
  = e^{i \left( {\phi}_{\vK'} - {\phi}_{\vK} \right)}
    \sin{\frac{{\theta}_{\vK'}}{2}} \sin{\frac{{\theta}_{\vK}}{2}}
  + \cos{\frac{{\theta}_{\vK'}}{2}} \cos{\frac{{\theta}_{\vK}}{2}}.
\end{equation}
The first two terms in \cref{eq:channels} lead to intravalley pairing,
and the third to intervalley pairing.
We analyze the possible states within mean field theory.
The BCS order parameter is
\begin{equation}
  {\chi}
  = v_0 \sumK \cc{g}_{\vK} \ev{c_{-\vK {\alpha}'} c_{\vK {\alpha}}},
\end{equation}
where the form of $g_{\vK}$ depends on the particular pairing channel.
The resulting Hamiltonian has the same form as the BCS Hamiltonian in
\cref{eq:induced}
but with an effective ${\Delta}_{\vK} = g_{\vK} {\cdot} {\chi}$.
The intravalley pairing has three symmetry channels,
with the couplings given by
$2 g_{\vK} = 1 +  \cos{{\theta}_{\vK}}$,
$\sqrt{2} e^{- i {\phi}_{\vK}} g_{\vK} = \sin{{\theta}_{\vK}}$
and $2 e^{- 2 i {\phi}_{\vK}} g_{\vK} = 1 - \cos{{\theta}_{\vK}}$.
For these channels, since
$\ev{c_{-\vK {\alpha}} c_{\vK {\alpha}}} = - \ev{c_{\vK {\alpha}} c_{-\vK {\alpha}}}$,
relabeling $\vK {\rightarrow} -\vK$ in the sum gives ${\chi} = 0$ %
\footnote{%
  For odd parity interactions, where $v \ofMK = -v \ofK$, the
  intravalley pairing is not excluded by symmetry.
  Specifically, repeating the calculation with this assumption,
  the intervalley terms fully cancel, and one obtains \cref{eq:channels}
  without the intervalley term on the third line.
}.
The intervalley pairing also has three symmetry channels:
$g_{\vK} = \sqrt{2}$,
$g_{\vK} = \sqrt{2} \cos{{\theta}_{\vK}}$,
and $g_{\vK} = \sqrt{2} \sin{{\theta}_{\vK}} \vc{\hat{k}}$.
Of the three,
the constant valued channel is dominant %
\footnote{%
  For example, using the values for \ce{WSe2},
  $\sin^2 {{\theta}_{\vK}} = 0.44$ and $\cos^2 {{\theta}_{\vK}} = 0.56$
  at the chemical potential.
}.
This is to be expected, as the local density-density interaction
leads to the largest pairing for electrons of opposite spins.
Since the intravalley processes have the same spin,
they are disfavored as compared to the intervalley pairing.

The key features of the intrinsic superconducting state
are identical to the proximally induced case when density-density
interactions dominate.
We restrict further analysis to that case,
and turn to the question of pair-breaking phenomena
induced either by optical or magnetic fields.
   \section{Optoelectronic coupling}

The non-interacting system displays valley selective optical excitations.
Light of a particular polarization only couples to one valley.
Since the superconducting state is
a coherent condensate admixing the two valleys,
we address whether pair-breaking displays similar valley selectivity.
In particular, we explore whether or not the two quasiparticles generated
by circularly polarized light, with total energy larger than
$E_g + {\Delta}_{\vK}$, occupy opposite valleys,
with one in the conduction band and the other in the valence band.

The optical excitations arise from the Berry curvature,
which acts as an effective angular momentum.
The electromagnetic potential $\vc{A}$,
with polarization vector $\boldsymbol{\epsilon}$,
is introduced using minimal coupling,
$H_{{\tau} {\s}}^{{\nu} {\nu}'} \ofK
{\rightarrow} H_{{\tau} {\s}}^{{\nu} {\nu}'} \of{\vK + e \vc{A}}$,
where, in the dipole approximation,
$\vc{A} = 2 \re{\boldsymbol{\epsilon} A_0 e^{- i {\omega} t}}$.
This yields a perturbed Hamiltonian
$H {\rightarrow} H + H^A$, where
$H^A = H' e^{- i {\omega} t} + H'^{\dagger} e^{i {\omega} t}$,
with
\begin{multline}
  H'
  = {\sum}_{\vK, {\tau}, {\s}}
    H'_{\tau}
    {d^-_{{\tau} {\s}}}^{\dagger} \ofK
    d^+_{{\tau} {\s}} \ofK \\
  - {\sum}_{\vK, {\tau}, {\s}}
    H'_{-{\tau}}
    {d^+_{{\tau} {\s}}}^{\dagger} \ofK
    d^-_{{\tau} {\s}} \ofK,
\end{multline}
and
$H'_{\tau}
= a t e A_0
\left( {\tau} \vc{\hat{x}} + i \vc{\hat{y}} \right) {\cdot} \boldsymbol{\epsilon}$.
The transition rate is proportional to the modulus-squared
of the optical matrix elements,
$\vc{P}_{{\tau} {\s}}^{n n'} \ofK$,
defined by
\begin{equation}
  H^A
  = {\sum}_{\substack{\vK, {\tau}, {\s} \\ n, n'}}
    \frac{e A_0}{m_0}
    \boldsymbol{\epsilon} {\cdot} \vc{P}_{{\tau} {\s}}^{n n'} \ofK
    {c_{{\tau} {\s}}^n}^{\dagger} \ofK
    c_{{\tau} {\s}}^{n'} \ofK.
\end{equation}
For circularly polarized light, in the absence of superconductivity,
$\boldsymbol{\epsilon}_{\pm} = \left( \vc{\hat{x}} {\pm} i \vc{\hat{y}} \right) / \sqrt{2}$ and
\begin{equation}
  \label{eq:optical}
  \boldsymbol{\epsilon}_{\pm} {\cdot} \vc{P}_{{\tau} {\s}}^{+ -} \of{\vK}
  = {\mp} {\tau} \sqrt{2} a t m_0
    e^{{\pm} i {\phi}}
    \sin^2 {\frac{\fnTheta{{\mp} {\tau}}}{2}}.
\end{equation}

\begin{figure}
  \includegraphics[width=\columnwidth]{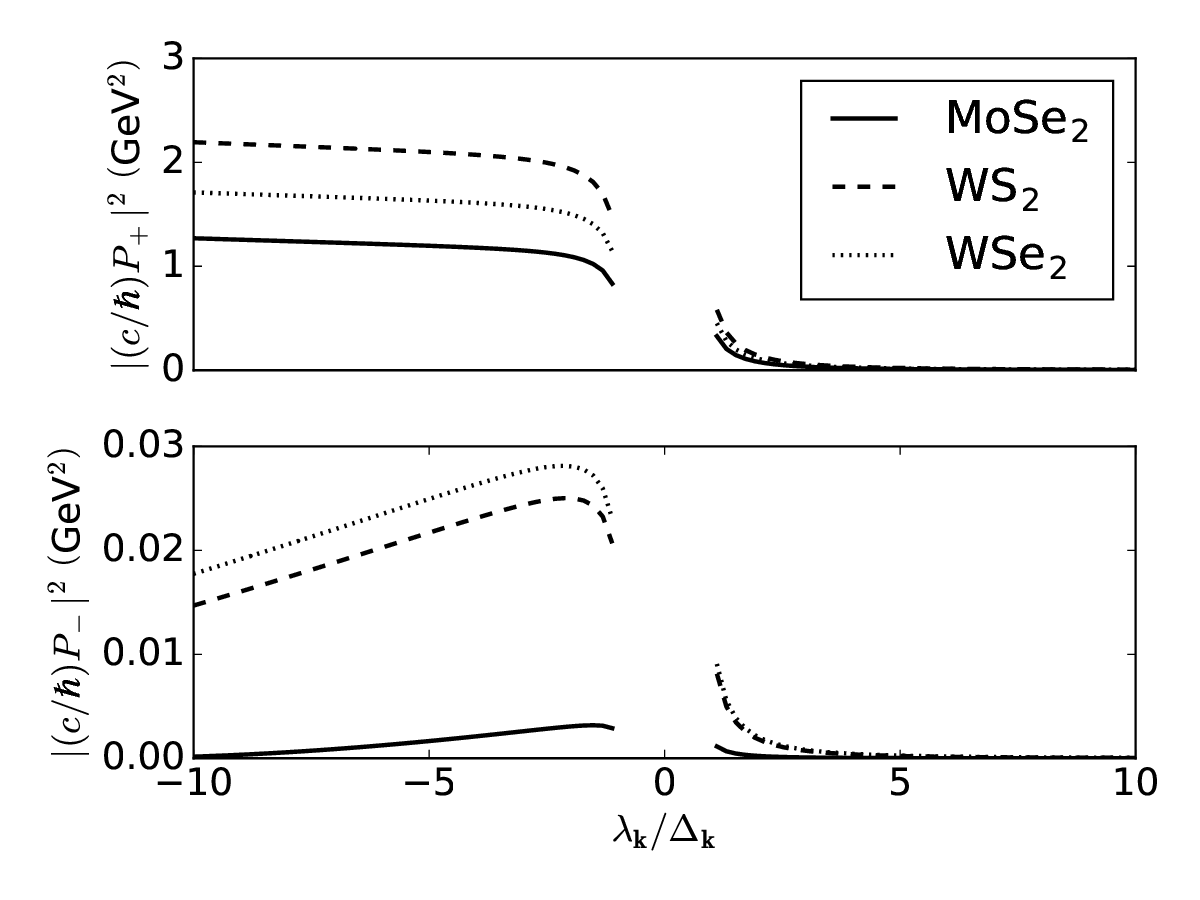}
  \caption{%
    Optical transition rate matrix elements
    $\left| P_{\pm} \right|^2$
    in the superconducting phase
    as a function of the ratio of the quasiparticle energy
    ${\lambda}_{\vK}$ to the superconducting gap ${\Delta}_{\vK}$.
    Material parameters for \ce{MoSe2}, \ce{WS2}, and \ce{WSe2}
    are given in~\cite{PhysRevLett.108.196802}
    and a gap of ${\Delta}_{\vK} = \SI{7.5}{\milli\electronvolt}$
    is chosen for illustrative purposes.
    The order-of-magnitude contrast between
    $\left|P_+\right|^2$ and $\left|P_-\right|^2$
    causes the optical-valley selectivity.
  }\label{fig:optical}
\end{figure}

\begin{figure}
  \includegraphics[width=\columnwidth]{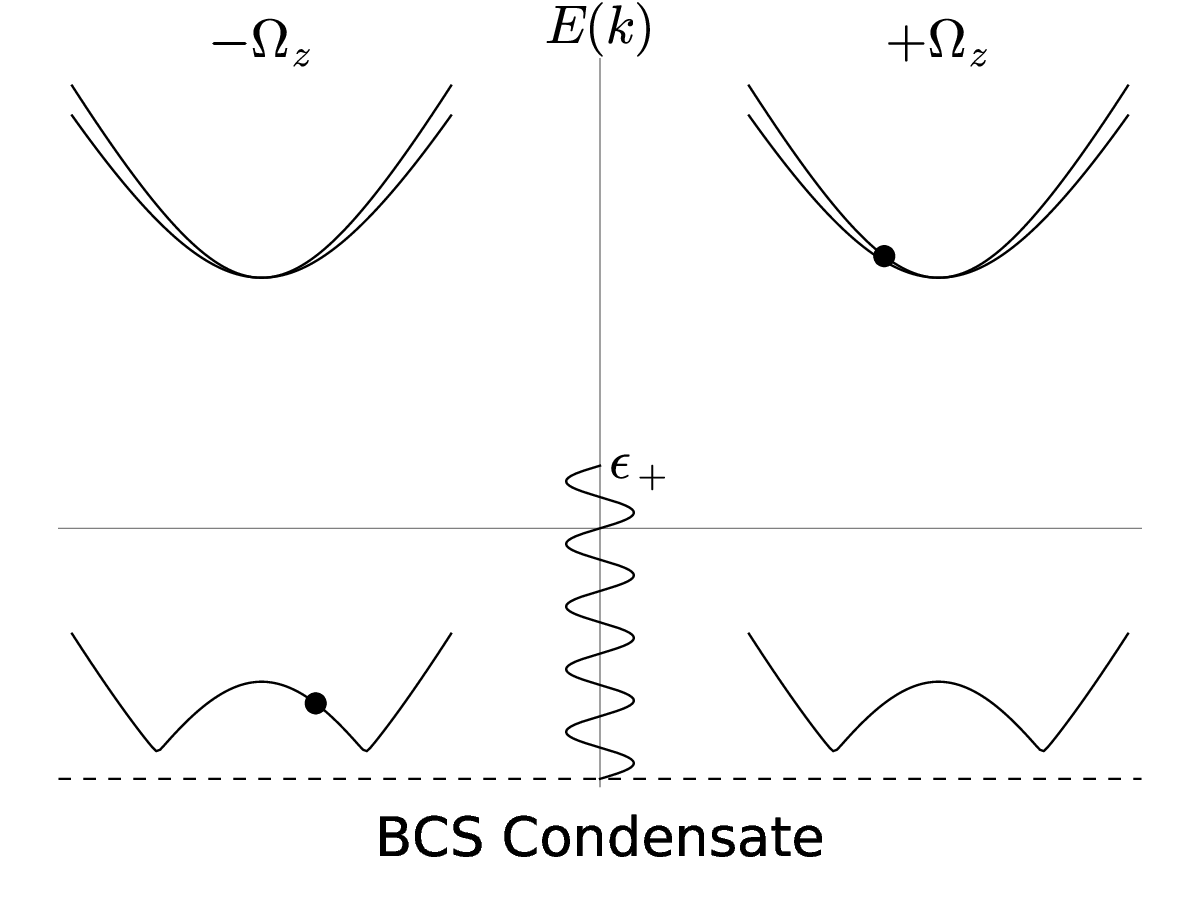}
  \caption{%
    Pair-breaking by right circularly polarized light
    leads to an electron in the conduction band of the right valley
    and a partner in the valence band of the left valley.
    The valleys interchange for left circularly polarized light.
  }\label{fig:optical-excitation}
\end{figure}

The transition rate matrix elements
for optical excitations from the BCS ground state
are given by \cref{eq:optical}
multiplied by a coherence factor $\sin {{\beta}_{\vK}}$.
Since $\fnTheta{-} - \fnTheta{+} = {\tau} {\pi}$,
switching either the valley or polarization transforms
$\sin {\rightarrow} \cos$ in \cref{eq:optical}, giving matrix elements
$\abs{P_{\pm}} = \abs{\boldsymbol{\epsilon}_{\pm} {\cdot} \vc{P}_{++}^{+ -} \ofK \sin {{\beta}_{\vK}}}$
corresponding to matching ($P_+$) or mismatching ($P_-$)
polarization-valley indexes.
For a given valley, a chosen polarization of light couples more strongly
than the other, as is evident comparing $\abs{P_+}^2$ to $\abs{P_-}^2$
and shown in \cref{fig:optical}.
For incident light with energy $E_g + \abs{{\lambda}_{\vK}}$,
right circularly polarized light ($+$) has a higher probability
of promoting a quasiparticle to the right conduction band,
as reflected in the larger matrix element $\abs{P_+}^2 {\gg} \abs{P_-}^2$.
As depicted in \cref{fig:optical-excitation},
the partner of the Cooper pair is in the valence band in the opposite valley.
The other valley has the opposite dependence on polarization.

This key new result opens the door for valley control of excitations
from a coherent ground state.
For example, the two quasiparticles have the same charge and Berry curvature
(see below).
In the presence of an electric filed,
they both acquire the same transverse anomalous velocity.
Thus, in contrast to the response in the normal state,
an anomalous Hall effect is anticipated
with no accompanying spin current.

   \section{Berry curvature}

The Berry curvature in the non-interacting crystal
for left and right circularly polarized
($\boldsymbol{\epsilon}_{\pm}$) optical excitations for a given $\vK$
is ${\pm} 2 {\Omega}_{+ {\uparrow}}^+ \of{k}$, where
\begin{subequations}
  \begin{align}
    {\Omega}_{{\tau} {\s}}^n \of{k}
    & = \vc{\hat{z}} {\cdot} \vc{{\Omega}}_{{\tau} {\s}}^n \ofK, \\
    & = - n {\tau}
        \left[ \frac{1}{2 k} \pderiv{}{k} \fnTheta{n} \right]
        \sin{\fnTheta{n}}, \\
    & = - n {\tau}
        \frac{2 {\left( a t \right)}^2 \left( E_g - \tau s E_{\text{soc}} \right)}
        {{\left[{\left( 2 a t k \right)}^2
      + {\left( E_g - \tau s E_{\text{soc}} \right)}^2 \right]}^{3/2}}.
  \end{align}
\end{subequations}

The BCS ground state %
\footnote{%
  Note that the full ground state
  also contains the two lower filled bands,
  but those contribute zero net Berry curvature and may be ignored
  in this section and the next.}
is
\begin{subequations}
  \begin{align}
    \Ket{{\Omega}}
    & = {\prod}_{\vK} \csc{{\beta}_{\vK}} {\gamma}_{\vK {\uparrow}} {\gamma}_{-\vK {\downarrow}} \Ket{0}, \\
    & = {\prod}_{\vK} \left( \cos{{\beta}_{\vK}} - \sin{{\beta}_{\vK}}
        c_{\vK {\uparrow}}^{\dagger} c_{-\vK {\downarrow}}^{\dagger} \right) \Ket{0}.
  \end{align}
\end{subequations}
This superconducting state is built up
from the quasiparticle eigenstates,
$\Ket{\vK}
= \csc{{\beta}_{\vK}} {\gamma}_{\vK {\uparrow}} {\gamma}_{-\vK {\downarrow}} \Ket{0}$,
of the $\vK$-dependent Hamiltonian
${\lambda}_{\vK} \left( {\gamma}_{\vK {\uparrow}}^{\dagger} {\gamma}_{\vK {\uparrow}}
+ {\gamma}_{-\vK {\downarrow}}^{\dagger} {\gamma}_{-\vK {\downarrow}} \right)$.
The $z$-component of the Berry curvature of
the correlated state is zero,
\begin{equation}
  \vc{\hat{z}} {\cdot} i {\nabla}_{\vK} {\times}
  \Braket{\vK | {\nabla}_{\vK} | \vK}
  = {\Omega}_{+ {\uparrow}}^- \of{k} + {\Omega}_{- {\downarrow}}^- \of{-k} = 0.
\end{equation}
A single optically excited state in the left valley
for a given $\vK$ is
${c_{+ {\uparrow}}^+}^{\dagger} \ofK c_{+ {\uparrow}}^- \Ket{\vK}$,
which has a Berry curvature
$+2 \sin^6 {{\beta}_{\vK}} {\Omega}_{+ {\uparrow}}^+ \of{k}$.
The corresponding excitation in the right valley
has a Berry curvature of the same magnitude but opposite sign.
   \section{In-plane magnetic field and scalar disorder}

In this section we discuss the effects of in-plane magnetic fields
and non-magnetic impurities on the superconducting state.
We consider the lightly hole-doped monolayer TMDs in the regime
where the Fermi level crosses the upper valence bands
and is well separated from the lower valence bands.
In this regime, the system is a spin-valley locking
system with the spin-opposite Fermi pocket at each valley.
Without loss of generality, we adopt a simplified model taking
into account the valence bands only.
In a quasi two-dimensional (2D) system,
an in-plane magnetic field couples to quasiparticles through
spin paramagnetism with negligible orbital interactions.
Applying a uniform in-plane magnetic field in the $x$ direction
$\vc{B} = \left( B, 0, 0 \right)$,
the system is described by the Hamiltonian (${\hbar} = k_B = c = 1$)
\begin{equation}
  \mathcal{H}_{\tau} \ofK
  = - \frac{k^2}{2 m} - {\mu} + {\tau} E_{\text{soc}} \hat{s}_z + {\mu}_B B \hat{s}_x,
\end{equation}
which is acting on the valley-spin basis
${\phi}_{\tau} \ofK = {\left( c_{{\tau} {\uparrow}} \ofK, c_{{\tau} {\downarrow}} \ofK \right)}^T$,
where ${\mu}$ is the chemical potential, ${\tau} = {\pm}$ is the valley
index, $\hat{s}_i$ are Pauli matrices operating in spin space,
and ${\mu}_B$ is the Bohr magneton.
The dispersion relations of the upper and lower valence bands
have been approximated by a quadratic form with an effective mass
$m = E_g / \left( 2 a^2 t^2 \right)$, where
$E_g$ is the large energy gap between the conduction and valence bands,
$E_g {\gg} E_{\text{soc}}$,
$a$ and $t$ are defined under \cref{eq:hamiltonian},
and the momentum $\vK = \left( k_x, k_y \right)$
is measured from the corresponding valley center with $k = \abs{\vK}$.
Note that in this section we use $\vK$ to represent momentum
measured from the corresponding valley center and $\vc{p}$ to
represent momentum measured from the Brillouin zone (BZ) center.
We use the notation that $c_{{\tau} \s}^{\dagger} \ofK$ ($c_{{\tau} \s} \ofK$)
creates (annihilates) a quasiparticle with momentum $\vK$
and spin $\s$ in the valley ${\tau}$, and $c_s^{\dagger} \of{\vc{p}}$
($c_{\s} \of{\vc{p}}$) creates (annihilates) a quasiparticle with
momentum $\vc{p}$ and spin $\s$.

The Hamiltonian $\mathcal{H}_{\tau} \ofK$ has the spectrum
\begin{equation}
  E_{{\tau}, u / l} \of{k}
  = - \frac{k^2}{2 m} -{\mu} {\pm} \sqrt{E_{\text{soc}}^2 + {\left( {\mu}_B B \right)}^2},
\end{equation}
with $u$ for the upper ($+$) and $l$ for the lower ($-$) band at each valley,
and the eigenstates
${\varphi}_{\tau} \ofK = {\left( c_{{\tau} u} \ofK, c_{{\tau} l} \ofK \right)}^T$,
where $c_{{\tau} u}$ and $c_{{\tau} l}$ correspond to the quasiparticles
in the band basis which is related to the spin basis through a field
and valley-dependent unitary transformation $U_{\tau} \of{b}$:
${\varphi}_{\tau} \ofK = U_{\tau} \of{b} {\phi}_{\tau} \ofK$,
where $b = {\mu}_B B / {E_{\text{soc}}}$
is the dimensionless magnetic field.
Applying a uniform in-plane magnetic field shifts both the
upper (lower) valence bands at the two valleys by the same amount
(but the opposite amount between the upper and lower band at each valley),
so that the perfect nesting condition between
the Fermi pockets at the two valleys remains.
Meanwhile, the quasiparticle spin acquires a finite in-plane component,
i.e., deviating from ${\pm} z$ direction.
Explicitly, we have
$\Braket{{\tau}, u | \hat{s}_z | {\tau}, u} = - \Braket{{\tau}, l | \hat{s}_z | {\tau}, l}
= \left( {\tau} / 2 \right) / \sqrt{1 + b^2}$,
and
$\Braket{{\tau}, u | \hat{s}_x | {\tau}, u} = - \Braket{{\tau}, l | \hat{s}_x | {\tau}, l}
= \left( b / 2 \right) / \sqrt{1 + b^2}$.
Therefore, at both valleys,
the quasiparticle spin tilts towards the field direction
in the upper valence bands and tilts against the field direction in
the lower valence bands.
The change of quasiparticle spin orientation induced by the in-plane field
modifies the internal structure of the Cooper pair
and affects the pairing strength as shown below.

To evaluate the effect of the magnetic field on the superconductivity
we follow the procedure used in \cref{s:superconductivity}.
A local attractive density-density
interaction with pairing strength $v_0$ can be written as:
$\mathcal{H}_{\text{int}} = - v_0 {\int} \dif^2 \vc{r} {\rho} \of{\vc{r}} {\rho} \of{\vc{r}}$
with the quasiparticle density
${\rho} \of{\vc{r}} = {\sum}_{\s} c_{\s}^{\dagger} \of{\vc{r}} c_{\s} \of{\vc{r}}$,
where $c_{\s}^{\dagger} \of{\vc{r}}$ ($c_{\s} \of{\vc{r}}$) is the
Fourier transform of $c_{\s}^{\dagger} \of{\vc{p}}$ ($c_{\s} \of{\vc{p}}$).
Transforming to momentum space and projecting onto the upper valence
bands, the pairing Hamiltonian has the form
\begin{equation}
  \mathcal{H}_p
  = - v' \of{b} {\sum}_{\vK, \vK'}
    c_+^{\dagger} \ofK c_-^{\dagger} \ofMK
    c_- \of{-\vK'} c_+ \of{\vK'},
\end{equation}
where we have ignored the upper-band subscript $u$ in the operators
$c_{{\tau} u}^{\dagger}$ and $c_{{\tau} u}$.
The effective pairing strength in the presence of in-plane magnetic field is
$v' \of{b} = v_0 / \left( 1 + b^2 \right)$.
The Hamiltonian $\mathcal{H}_p$ describes an intervalley pairing
with a pairing strength $v'$ suppressed by the in-plane field.
At zero field, $v' = v_0$,
$c_+^{\dagger} = c_{+ {\uparrow}}^{\dagger}$, and $c_-^{\dagger} = c_{- {\downarrow}}^{\dagger}$,
so the pairing occurs between opposite spins.
At finite fields, $v' < v_0$ and the quasiparticle at valley
$+$ ($-$), represented by $c_+^{\dagger}$ ($c_-^{\dagger}$),
has its up (down) spin tilted towards the field direction.
As a result, the intervalley pairing contains equal-spin pairing components in
the presence of in-plane field.

The mean-field Hamiltonian, using the Nambu-valley basis
 ${\Psi}_{\vK} = {\left( c_+ \ofK, c_- \ofK, c_-^{\dagger} \of{-\vK}, - c_+^{\dagger} \of{-\vK} \right)}^T$,
takes the form
\begin{equation}
  \mathcal{H}_{\text{MF}} \ofK
  = {\xi}_{\vK} \hat{{\eta}}_z - {\Delta} \hat{{\eta}}_x,
\end{equation}
where $\hat{{\eta}}_i$ are Pauli matrices acting on Nambu (particle-hole) space,
${\xi}_{\vK}
= - k^2 / 2 m - {\mu} + \sqrt{E_{\text{soc}}^2 + {\left( {\mu}_B B \right)}^2}$,
and the mean field
${\Delta} = v' \of{b} {\sum}_{\vK'} \braket{c_- \of{-\vK'} c_+ \of{\vK'}}$
describes an intervalley pairing field, which we choose to be real
for convenience.

In a conventional 2D superconductor with a spin-degenerate Fermi surface,
the application of an in-plane magnetic field induces an energy splitting
between opposite-spin bands.
This energy mismatch between opposite spins creates
a pair-breaking effect in the clean system characterized
by the pair-breaking equation for temperature $T~{{\leq}}~T_c^0$
\cite{Maki01061964},
\begin{multline}
  \label{eq:conventional_pair_breaking_eq}
  \ln{\frac{T_c^0}{T}} \\
  = \frac{1}{2} \left[ {\psi} \of{\frac{1}{2} + \frac{i {\mu}_B B_c}{2 {\pi} T}}
  + {\psi} \of{\frac{1}{2} - \frac{i {\mu}_B B_c}{2 {\pi} T}} \right] -  {\psi} \of{\frac{1}{2}},
\end{multline}
where $T_c^0$ is the transition temperature at zero field in
the clean system and ${\psi} \of{z}$ is the digamma function.
This equation determines the critical field $B_c$
that destroys the superconducting state at temperature $T~{{\leq}}~T_c^0$
from spin paramagnetism.
Furthermore, the scattering from non-magnetic impurities
does not alter this pair-breaking
\cref{eq:conventional_pair_breaking_eq} such that the
critical field remains the same regardless of the presence of scalar
disorders
\cite{Maki01061964}.

Unlike the conventional 2D superconductors,
in our system the two single-spin Fermi pockets at different valleys
remain perfectly nested without the energy mismatch
caused by spin paramagnetism,
and the spins at the two pockets are no longer opposite with equal-spin
components induced by the field.
These two differences give rise to new features
in the spin-valley locking system from the effects of in-plane magnetic fields.
First, in the clean limit, the presence of in-plane
magnetic fields does not lead to a pair-breaking effect for the lack
of energy mismatch, but mildly suppresses the transition temperature
through the weakening of the pairing strength.
The suppressed transition
temperature $T_c'$ is related to the zero-field transition temperature $T_c^0$ as
$T_c' = T_c^0 \exp{\left( - b^2 / v_0 N_F \right)}$
in the mean-field theory,
where $N_F$ is the density of states at the Fermi level.
Second, the superconducting state is no longer
immune to the scalar disorder, since non-magnetic disorder potential
can cause intervalley scattering due to the field-induced parallel-spin
components on the two pockets.
This interplay between the in-plane magnetic field
and the scalar disorder leads to a pair-breaking effect.

In the presence of dilute randomly-distributed non-magnetic impurities,
the Hamiltonian for short-range impurity potential is given by
\begin{equation}
  \mathcal{H}_{\text{imp}}
  = {\sum}_j {\int} \dif^2 \vc{r} U_0 {\delta} \of{\vc{r} - \vc{R}_j} {\rho} \of{\vc{r}},
\end{equation}
where $\vc{R}_j$ is the position of the $j$th impurity and
$U_0$ is the disorder strength.
Transforming to momentum space
and projecting onto the upper valence bands, $\mathcal{H}_{\text{imp}}$
can be written using the Nambu-valley basis ${\Psi}_{\vK}$ as
\begin{equation}
  \mathcal{H}_{\text{imp}}
  = {\sum}_{{\vK}_1, {\vK}_2} {\sum}_j
    e^{i \left( {\vK}_1 - {\vK}_2 \right) {\cdot} \vc{R}_j}
    {\Psi}_{{\vK}_1}^{\dagger} \hat{U} {\Psi}_{{\vK}_2},
\end{equation}
with the disorder scattering vertex $\hat{U}$ taking the form
\begin{equation}
  \hat{U}
  = U_0 \hat{{\eta}}_z + U_0 \frac{b}{\sqrt{1 + b^2}} \hat{{\tau}}_x,
\end{equation}
where $\hat{{\tau}}_i$ are Pauli matrices operating in valley space.
The first term in $\hat{U}$ corresponds to intravalley scattering
and the second term corresponds to intervalley scattering.
Note that we have ignored the factors
$e^{i \left( {\pm} 2 \vc{K} \right) {\cdot} \vc{R}_j }$
in the intervalley terms because
$e^{i \left( 2 \vc{K} \right) {\cdot} \vc{R}_j}$
and $e^{i \left( - 2 \vc{K} \right) {\cdot} \vc{R}_j}$
will appear in pair and cancel each other
in the diagrammatical calculation of self energy.

The self energy due to impurity scattering
after averaging over randomly-distributed impurity configurations,
in the first-order Born approximation, is obtained as
\cite{AbrikosovGorkov1961,maki1969superconductivity}
\begin{equation}
  \hat{{\Sigma}} \of {\vK, i {\omega}_n}
  = n_{\text{imp}}
    {\int} \frac{\dif^2 \vK'}{{\left( 2{\pi} \right)}^2}
    \hat{U} \hat{\mathcal{G}}_0 \of{\vK', i {\omega}_n} \hat{U},
\end{equation}
where $n_{\text{imp}}$ is the impurity concentration
and $\hat{\mathcal{G}}_0$
is the Green's function matrix of the clean system,
$\hat{\mathcal{G}}_0 \of{\vK', i {\omega}_n}
= {\left( i {\omega}_n - {\xi}_{\vK'} \hat{{\eta}}_z + {\Delta} \hat{{\eta}}_x \right )}^{-1}$,
with the Matsubara frequencies
${\omega}_n = \left( 2 n + 1 \right) {\pi} T$.
After integrating out ${\xi}$ in the self-energy,
the disorder renormalized Green's function matrix
$\hat{\mathcal{G}} = {\left( \hat{\mathcal{G}}_0^{-1} - \hat{{\Sigma}} \right)}^{-1}$
can be parametrized as
\begin{equation}
  \label{eq:renormalized_g}
  \hat{\mathcal{G}} \of{\vK, i {\omega}_n}
  = {\left[ i \tilde{{\omega}}_n - {\xi}_{\vK} \hat{{\eta}}_z
    + \tilde{{\Delta}} \hat{{\eta}}_x + i F \of{{\omega}_n} \hat{{\eta}}_z \hat{{\tau}}_x \right]}^{-1}, \!
\end{equation}
where the quantities
$\tilde{{\omega}}_n$, $\tilde{{\Delta}}$, and $F \of{{\omega}_n}$ have the definitions
\begin{equation}
  \tilde{{\omega}}_n
  = {\omega}_n + \left( \frac{1}{2 {\tau}_1} + \frac{1}{2 {\tau}_2} \right)
    \frac{{\omega}_n}{\sqrt{{\omega}_n^2 + {\Delta}^2}},
\end{equation}
\begin{equation}
  \tilde{{\Delta}}
  = {\Delta} + \left( \frac{1}{2 {\tau}_1} - \frac{1}{2 {\tau}_2} \right)
    \frac{{\Delta}}{\sqrt{{\omega}_n^2 + {\Delta}^2}},
\end{equation}
\begin{equation}
  F \of{{\omega}_n}
  = \frac{1}{2 {\tau}_1} \frac{{\omega}_n}{\sqrt{{\omega}_n^2 + {\Delta}^2}} \frac{2b}{\sqrt{1 + b^2}}.
\end{equation}
Here, $1 / {\tau}_1$ and $1 / {\tau}_2$ are the collision rates corresponding
to the disorder-induced intravalley and intervalley scattering, respectively,
with the expressions
\begin{equation}
  \frac{1}{{\tau}_1} = 2 U_0^2 n_{\text{imp}} {\pi} N_F, \:
  \frac{1}{{\tau}_2} = \frac{1}{{\tau}_1} \frac{b^2}{1 + b^2}.
\end{equation}

In the superconducting state, the self-consistency equation for the
order parameter is given by
\begin{equation}
  {\Delta}
  = \frac{1}{4} v' T
    {\sum}_n {\int} \frac{\dif^2 \vK}{{\left( 2 {\pi} \right)}^2}
    \Tr{\left[ \hat{{\eta}}_x \hat{\mathcal{G}} \of{\vK, i {\omega}_n} \right]},
\end{equation}
where $\Tr \left[ {\ldots} \right]$ is the trace of the argument.
Explicitly, from \cref{eq:renormalized_g}, it has the form
\begin{equation}
  \label{eq:explicit_self_consistency_eq}
  {\Delta} = v' {\pi} N_F T {\sum}_n \frac{\tilde{{\Delta}}}{\sqrt{\tilde{{\omega}}_n^2 + \tilde{{\Delta}}^2}}.
\end{equation}
Linearizing the self-consistency \cref{eq:explicit_self_consistency_eq}
near the critical field $B_c$, we obtain the pair-breaking equation
due to the interplay between the in-plane field and the scalar disorder,
\begin{equation}
  \label{eq:pair_breaking_eq}
  \ln{\frac{T_c' \of{b_c}}{T}}
  = {\psi} \of{\frac{1}{2} + \frac{{\delta}_c}{2 {\pi} T}} - {\psi} \of{\frac{1}{2}},
\end{equation}
where $T_c' \of{b_c} = T_c^0 \exp{\left( - b_c^2 / v_0 N_F \right)}$
is the transition temperature in the clean system in the presence of the field
$b_c = {\mu}_B B_c / E_{\text{soc}}$,
and the pair-breaking parameter
\begin{equation}
  {\delta}_c = \frac{1}{{\tau}_2} \biggl{\rvert}_{b_c}
      = \frac{1}{{\tau}_1} \frac{b_c^2}{1 + b_c^2}
\end{equation}
arises from the valley-flip scattering process.
\Cref{eq:pair_breaking_eq} determines the in-plane critical
field $B_c \of{T}$ at temperature $T~{{\leq}}~T_c'$.
For $b_c~{{\ll}}~1$, the pair-breaking parameter takes the simple form
${\delta}_c~{{\approx}}~{\tau}_1^{-1} {\left( {\mu}_B B_c / E_{\text{soc}} \right)}^2$.

As $T~{{\rightarrow}}~0$, the pair-breaking \cref{eq:pair_breaking_eq}
can be approximated by the asymptotic expansion of the digamma function,
which leads to
$2 {\pi} T_c' \exp{\left[ {\psi} \of{1 / 2} \right]}
= \left( b_c^2 / {\tau}_1 \right) / \left( 1 + b_c^2 \right)$.
At finite disorder concentration, ${\tau}_1^{-1}~{{\neq}}~0$,
when $b_c~{{\ll}}~1$ with $T_c'~{{\approx}}~T_c^0$,
the critical field at zero temperature is approximated as
\begin{equation}
  {\mu}_B B_c \biggl{\rvert}_{T~{{\rightarrow}}~0}~{{\approx}}~E_{\text{soc}}
    {\left[ 2{\pi} e^{{\psi} \of{1/2}} k_B T_c^0 {\tau}_1 / {\hbar} \right]}^{1/2},
\end{equation}
where we have put back the Boltzmann constant $k_{B}$ and
the reduced Planck constant ${\hbar}$.
The large spin-orbit interaction $E_{\text{soc}}$
($\sim$\SIrange[range-phrase=--, range-units=single]{150}{500}{{\milli}\electronvolt})
in monolayer TMDs indicates that the in-plane
critical field $B_c$ is significantly enhanced, well beyond the
Pauli limit.
   \section{Conclusions}

In this letter, we report on the nature of the superconducting state
of hole-doped TMDs.
Remarkably, the correlated state inherits
the valley contrasting phenomena of the non-interacting state.
While the magnitude is smaller, pair-breaking produces quasiparticles
that have the same Berry curvature, and hence the same anomalous velocity.
Thus one predicts an anomalous Hall response unlike the valley Hall response
observed in \ce{MoSe2}.

Spin-valley locking leads to large critical magnetic fields.
A similar phenomena was recently reported in heavily hole-doped
(beyond the spin-split gap) \ce{NbSe2}
\cite{Xi2016}.
In the new regime, where only one band per valley intersects
the chemical potential, no pair-breaking occurs
for in-plane fields unless disorder is present.

While systematic synthesis and characterization of hole-doped systems
is still in its early stages, the fact that other two-dimensional compounds
and their bulk counterparts are known to be superconducting
\cite{%
  PhysRevB.88.054515%
}
provides impetus to explore this novel phenomena.
   \begin{acknowledgments}
  The software developed %
\footnote{%
  Related software and source code at \\
  \url{https://evansosenko.com/dichalcogenides}
}
  and used~\cite{Hunter:2007} for this work
  and the included figures is available freely online.
  We thank Michael Phillips for useful discussions.
  We acknowledge the support of the Army Research Office through the grant
  ARO W911NF1510079.
\end{acknowledgments}
 
  % Must keep on single line or latexpand will fail.
 

\begin{thebibliography}{44}%
\makeatletter
\providecommand \@ifxundefined [1]{%
 \@ifx{#1\undefined}
}%
\providecommand \@ifnum [1]{%
 \ifnum #1\expandafter \@firstoftwo
 \else \expandafter \@secondoftwo
 \fi
}%
\providecommand \@ifx [1]{%
 \ifx #1\expandafter \@firstoftwo
 \else \expandafter \@secondoftwo
 \fi
}%
\providecommand \natexlab [1]{#1}%
\providecommand \enquote  [1]{``#1''}%
\providecommand \bibnamefont  [1]{#1}%
\providecommand \bibfnamefont [1]{#1}%
\providecommand \citenamefont [1]{#1}%
\providecommand \href@noop [0]{\@secondoftwo}%
\providecommand \href [0]{\begingroup \@sanitize@url \@href}%
\providecommand \@href[1]{\@@startlink{#1}\@@href}%
\providecommand \@@href[1]{\endgroup#1\@@endlink}%
\providecommand \@sanitize@url [0]{\catcode `\\12\catcode `\$12\catcode
  `\&12\catcode `\#12\catcode `\^12\catcode `\_12\catcode `\%12\relax}%
\providecommand \@@startlink[1]{}%
\providecommand \@@endlink[0]{}%
\providecommand \url  [0]{\begingroup\@sanitize@url \@url }%
\providecommand \@url [1]{\endgroup\@href {#1}{\urlprefix }}%
\providecommand \urlprefix  [0]{URL }%
\providecommand \Eprint [0]{\href }%
\providecommand \doibase [0]{http://dx.doi.org/}%
\providecommand \selectlanguage [0]{\@gobble}%
\providecommand \bibinfo  [0]{\@secondoftwo}%
\providecommand \bibfield  [0]{\@secondoftwo}%
\providecommand \translation [1]{[#1]}%
\providecommand \BibitemOpen [0]{}%
\providecommand \bibitemStop [0]{}%
\providecommand \bibitemNoStop [0]{.\EOS\space}%
\providecommand \EOS [0]{\spacefactor3000\relax}%
\providecommand \BibitemShut  [1]{\csname bibitem#1\endcsname}%
\let\auto@bib@innerbib\@empty
%</preamble>
\bibitem [{\citenamefont {Haldane}(1988)}]{PhysRevLett.61.2015}%
  \BibitemOpen
  \bibfield  {author} {\bibinfo {author} {\bibfnamefont {F.~D.~M.}\
  \bibnamefont {Haldane}},\ }\href {\doibase 10.1103/PhysRevLett.61.2015}
  {\bibfield  {journal} {\bibinfo  {journal} {Phys. Rev. Lett.}\ }\textbf
  {\bibinfo {volume} {61}},\ \bibinfo {pages} {2015} (\bibinfo {year}
  {1988})}\BibitemShut {NoStop}%
\bibitem [{\citenamefont {Kane}\ and\ \citenamefont
  {Mele}(2005)}]{PhysRevLett.95.226801}%
  \BibitemOpen
  \bibfield  {author} {\bibinfo {author} {\bibfnamefont {C.~L.}\ \bibnamefont
  {Kane}}\ and\ \bibinfo {author} {\bibfnamefont {E.~J.}\ \bibnamefont
  {Mele}},\ }\href {\doibase 10.1103/PhysRevLett.95.226801} {\bibfield
  {journal} {\bibinfo  {journal} {Phys. Rev. Lett.}\ }\textbf {\bibinfo
  {volume} {95}},\ \bibinfo {pages} {226801} (\bibinfo {year}
  {2005})}\BibitemShut {NoStop}%
\bibitem [{\citenamefont {Bernevig}\ and\ \citenamefont
  {Zhang}(2006)}]{PhysRevLett.96.106802}%
  \BibitemOpen
  \bibfield  {author} {\bibinfo {author} {\bibfnamefont {B.~A.}\ \bibnamefont
  {Bernevig}}\ and\ \bibinfo {author} {\bibfnamefont {S.-C.}\ \bibnamefont
  {Zhang}},\ }\href {\doibase 10.1103/PhysRevLett.96.106802} {\bibfield
  {journal} {\bibinfo  {journal} {Phys. Rev. Lett.}\ }\textbf {\bibinfo
  {volume} {96}},\ \bibinfo {pages} {106802} (\bibinfo {year}
  {2006})}\BibitemShut {NoStop}%
\bibitem [{\citenamefont {K\"{o}nig}\ \emph {et~al.}(2007)\citenamefont
  {K\"{o}nig}, \citenamefont {Wiedmann}, \citenamefont {Br\"{u}ne},
  \citenamefont {Roth}, \citenamefont {Buhmann}, \citenamefont {Molenkamp},
  \citenamefont {Qi},\ and\ \citenamefont {Zhang}}]{Konig02112007}%
  \BibitemOpen
  \bibfield  {author} {\bibinfo {author} {\bibfnamefont {M.}~\bibnamefont
  {K\"{o}nig}}, \bibinfo {author} {\bibfnamefont {S.}~\bibnamefont {Wiedmann}},
  \bibinfo {author} {\bibfnamefont {C.}~\bibnamefont {Br\"{u}ne}}, \bibinfo
  {author} {\bibfnamefont {A.}~\bibnamefont {Roth}}, \bibinfo {author}
  {\bibfnamefont {H.}~\bibnamefont {Buhmann}}, \bibinfo {author} {\bibfnamefont
  {L.~W.}\ \bibnamefont {Molenkamp}}, \bibinfo {author} {\bibfnamefont {X.-L.}\
  \bibnamefont {Qi}}, \ and\ \bibinfo {author} {\bibfnamefont {S.-C.}\
  \bibnamefont {Zhang}},\ }\href {\doibase 10.1126/science.1148047} {\bibfield
  {journal} {\bibinfo  {journal} {Science}\ }\textbf {\bibinfo {volume}
  {318}},\ \bibinfo {pages} {766} (\bibinfo {year} {2007})}\BibitemShut
  {NoStop}%
\bibitem [{\citenamefont {Hasan}\ and\ \citenamefont
  {Kane}(2010)}]{RevModPhys.82.3045}%
  \BibitemOpen
  \bibfield  {author} {\bibinfo {author} {\bibfnamefont {M.~Z.}\ \bibnamefont
  {Hasan}}\ and\ \bibinfo {author} {\bibfnamefont {C.~L.}\ \bibnamefont
  {Kane}},\ }\href {\doibase 10.1103/RevModPhys.82.3045} {\bibfield  {journal}
  {\bibinfo  {journal} {Rev. Mod. Phys.}\ }\textbf {\bibinfo {volume} {82}},\
  \bibinfo {pages} {3045} (\bibinfo {year} {2010})}\BibitemShut {NoStop}%
\bibitem [{\citenamefont {Qi}\ and\ \citenamefont
  {Zhang}(2011)}]{RevModPhys.83.1057}%
  \BibitemOpen
  \bibfield  {author} {\bibinfo {author} {\bibfnamefont {X.-L.}\ \bibnamefont
  {Qi}}\ and\ \bibinfo {author} {\bibfnamefont {S.-C.}\ \bibnamefont {Zhang}},\
  }\href {\doibase 10.1103/RevModPhys.83.1057} {\bibfield  {journal} {\bibinfo
  {journal} {Rev. Mod. Phys.}\ }\textbf {\bibinfo {volume} {83}},\ \bibinfo
  {pages} {1057} (\bibinfo {year} {2011})}\BibitemShut {NoStop}%
\bibitem [{\citenamefont {Witczak-Krempa}\ \emph {et~al.}(2014)\citenamefont
  {Witczak-Krempa}, \citenamefont {Chen}, \citenamefont {Kim},\ and\
  \citenamefont {Balents}}]{doi:10.1146/annurev-conmatphys-020911-125138}%
  \BibitemOpen
  \bibfield  {author} {\bibinfo {author} {\bibfnamefont {W.}~\bibnamefont
  {Witczak-Krempa}}, \bibinfo {author} {\bibfnamefont {G.}~\bibnamefont
  {Chen}}, \bibinfo {author} {\bibfnamefont {Y.~B.}\ \bibnamefont {Kim}}, \
  and\ \bibinfo {author} {\bibfnamefont {L.}~\bibnamefont {Balents}},\ }\href
  {\doibase 10.1146/annurev-conmatphys-020911-125138} {\bibfield  {journal}
  {\bibinfo  {journal} {Annual Review of Condensed Matter Physics}\ }\textbf
  {\bibinfo {volume} {5}},\ \bibinfo {pages} {57} (\bibinfo {year}
  {2014})}\BibitemShut {NoStop}%
\bibitem [{\citenamefont {Radisavljevic}\ \emph {et~al.}(2011)\citenamefont
  {Radisavljevic}, \citenamefont {Radenovic}, \citenamefont {Brivio},
  \citenamefont {Giacometti},\ and\ \citenamefont {Kis}}]{RadisavljevicB.2011}%
  \BibitemOpen
  \bibfield  {author} {\bibinfo {author} {\bibfnamefont {B.}~\bibnamefont
  {Radisavljevic}}, \bibinfo {author} {\bibfnamefont {A.}~\bibnamefont
  {Radenovic}}, \bibinfo {author} {\bibfnamefont {J.}~\bibnamefont {Brivio}},
  \bibinfo {author} {\bibfnamefont {V.}~\bibnamefont {Giacometti}}, \ and\
  \bibinfo {author} {\bibfnamefont {A.}~\bibnamefont {Kis}},\ }\href {\doibase
  10.1038/nnano.2010.279} {\bibfield  {journal} {\bibinfo  {journal} {Nat
  Nano}\ }\textbf {\bibinfo {volume} {6}},\ \bibinfo {pages} {147} (\bibinfo
  {year} {2011})}\BibitemShut {NoStop}%
\bibitem [{\citenamefont {Zhu}\ \emph {et~al.}(2011)\citenamefont {Zhu},
  \citenamefont {Cheng},\ and\ \citenamefont
  {Schwingenschl\"ogl}}]{PhysRevB.84.153402}%
  \BibitemOpen
  \bibfield  {author} {\bibinfo {author} {\bibfnamefont {Z.~Y.}\ \bibnamefont
  {Zhu}}, \bibinfo {author} {\bibfnamefont {Y.~C.}\ \bibnamefont {Cheng}}, \
  and\ \bibinfo {author} {\bibfnamefont {U.}~\bibnamefont
  {Schwingenschl\"ogl}},\ }\href {\doibase 10.1103/PhysRevB.84.153402}
  {\bibfield  {journal} {\bibinfo  {journal} {Phys. Rev. B}\ }\textbf {\bibinfo
  {volume} {84}},\ \bibinfo {pages} {153402} (\bibinfo {year}
  {2011})}\BibitemShut {NoStop}%
\bibitem [{\citenamefont {Zhang}\ \emph {et~al.}(2012)\citenamefont {Zhang},
  \citenamefont {Ye}, \citenamefont {Matsuhashi},\ and\ \citenamefont
  {Iwasa}}]{doi:10.1021/nl2021575}%
  \BibitemOpen
  \bibfield  {author} {\bibinfo {author} {\bibfnamefont {Y.}~\bibnamefont
  {Zhang}}, \bibinfo {author} {\bibfnamefont {J.}~\bibnamefont {Ye}}, \bibinfo
  {author} {\bibfnamefont {Y.}~\bibnamefont {Matsuhashi}}, \ and\ \bibinfo
  {author} {\bibfnamefont {Y.}~\bibnamefont {Iwasa}},\ }\href {\doibase
  10.1021/nl2021575} {\bibfield  {journal} {\bibinfo  {journal} {Nano Letters}\
  }\textbf {\bibinfo {volume} {12}},\ \bibinfo {pages} {1136} (\bibinfo {year}
  {2012})},\ \bibinfo {note} {pMID: 22276648}\BibitemShut {NoStop}%
\bibitem [{\citenamefont {Wang}\ \emph {et~al.}(2012)\citenamefont {Wang},
  \citenamefont {Kalantar-Zadeh}, \citenamefont {Kis}, \citenamefont
  {Coleman},\ and\ \citenamefont {Strano}}]{Wang2012}%
  \BibitemOpen
  \bibfield  {author} {\bibinfo {author} {\bibfnamefont {Q.~H.}\ \bibnamefont
  {Wang}}, \bibinfo {author} {\bibfnamefont {K.}~\bibnamefont
  {Kalantar-Zadeh}}, \bibinfo {author} {\bibfnamefont {A.}~\bibnamefont {Kis}},
  \bibinfo {author} {\bibfnamefont {J.~N.}\ \bibnamefont {Coleman}}, \ and\
  \bibinfo {author} {\bibfnamefont {M.~S.}\ \bibnamefont {Strano}},\ }\href
  {\doibase 10.1038/nnano.2012.193} {\bibfield  {journal} {\bibinfo  {journal}
  {Nat Nano}\ }\textbf {\bibinfo {volume} {7}},\ \bibinfo {pages} {699}
  (\bibinfo {year} {2012})}\BibitemShut {NoStop}%
\bibitem [{\citenamefont {Ye}\ \emph {et~al.}(2012)\citenamefont {Ye},
  \citenamefont {Zhang}, \citenamefont {Akashi}, \citenamefont {Bahramy},
  \citenamefont {Arita},\ and\ \citenamefont {Iwasa}}]{Ye30112012}%
  \BibitemOpen
  \bibfield  {author} {\bibinfo {author} {\bibfnamefont {J.~T.}\ \bibnamefont
  {Ye}}, \bibinfo {author} {\bibfnamefont {Y.~J.}\ \bibnamefont {Zhang}},
  \bibinfo {author} {\bibfnamefont {R.}~\bibnamefont {Akashi}}, \bibinfo
  {author} {\bibfnamefont {M.~S.}\ \bibnamefont {Bahramy}}, \bibinfo {author}
  {\bibfnamefont {R.}~\bibnamefont {Arita}}, \ and\ \bibinfo {author}
  {\bibfnamefont {Y.}~\bibnamefont {Iwasa}},\ }\href {\doibase
  10.1126/science.1228006} {\bibfield  {journal} {\bibinfo  {journal}
  {Science}\ }\textbf {\bibinfo {volume} {338}},\ \bibinfo {pages} {1193}
  (\bibinfo {year} {2012})}\BibitemShut {NoStop}%
\bibitem [{\citenamefont {Bao}\ \emph {et~al.}(2013)\citenamefont {Bao},
  \citenamefont {Cai}, \citenamefont {Kim}, \citenamefont {Sridhara},\ and\
  \citenamefont {Fuhrer}}]{Bao2013}%
  \BibitemOpen
  \bibfield  {author} {\bibinfo {author} {\bibfnamefont {W.}~\bibnamefont
  {Bao}}, \bibinfo {author} {\bibfnamefont {X.}~\bibnamefont {Cai}}, \bibinfo
  {author} {\bibfnamefont {D.}~\bibnamefont {Kim}}, \bibinfo {author}
  {\bibfnamefont {K.}~\bibnamefont {Sridhara}}, \ and\ \bibinfo {author}
  {\bibfnamefont {M.~S.}\ \bibnamefont {Fuhrer}},\ }\href {\doibase
  10.1063/1.4789365} {\bibfield  {journal} {\bibinfo  {journal} {Applied
  Physics Letters}\ }\textbf {\bibinfo {volume} {102}},\ \bibinfo {eid}
  {042104} (\bibinfo {year} {2013}),\ 10.1063/1.4789365}\BibitemShut {NoStop}%
\bibitem [{\citenamefont {Zahid}\ \emph {et~al.}(2013)\citenamefont {Zahid},
  \citenamefont {Liu}, \citenamefont {Zhu}, \citenamefont {Wang},\ and\
  \citenamefont {Guo}}]{1.4804936}%
  \BibitemOpen
  \bibfield  {author} {\bibinfo {author} {\bibfnamefont {F.}~\bibnamefont
  {Zahid}}, \bibinfo {author} {\bibfnamefont {L.}~\bibnamefont {Liu}}, \bibinfo
  {author} {\bibfnamefont {Y.}~\bibnamefont {Zhu}}, \bibinfo {author}
  {\bibfnamefont {J.}~\bibnamefont {Wang}}, \ and\ \bibinfo {author}
  {\bibfnamefont {H.}~\bibnamefont {Guo}},\ }\href {\doibase 10.1063/1.4804936}
  {\bibfield  {journal} {\bibinfo  {journal} {AIP Advances}\ }\textbf {\bibinfo
  {volume} {3}},\ \bibinfo {eid} {052111} (\bibinfo {year} {2013}),\
  10.1063/1.4804936}\BibitemShut {NoStop}%
\bibitem [{\citenamefont {Cappelluti}\ \emph {et~al.}(2013)\citenamefont
  {Cappelluti}, \citenamefont {Rold\'an}, \citenamefont {Silva-Guill\'en},
  \citenamefont {Ordej\'on},\ and\ \citenamefont
  {Guinea}}]{PhysRevB.88.075409}%
  \BibitemOpen
  \bibfield  {author} {\bibinfo {author} {\bibfnamefont {E.}~\bibnamefont
  {Cappelluti}}, \bibinfo {author} {\bibfnamefont {R.}~\bibnamefont
  {Rold\'an}}, \bibinfo {author} {\bibfnamefont {J.~A.}\ \bibnamefont
  {Silva-Guill\'en}}, \bibinfo {author} {\bibfnamefont {P.}~\bibnamefont
  {Ordej\'on}}, \ and\ \bibinfo {author} {\bibfnamefont {F.}~\bibnamefont
  {Guinea}},\ }\href {\doibase 10.1103/PhysRevB.88.075409} {\bibfield
  {journal} {\bibinfo  {journal} {Phys. Rev. B}\ }\textbf {\bibinfo {volume}
  {88}},\ \bibinfo {pages} {075409} (\bibinfo {year} {2013})}\BibitemShut
  {NoStop}%
\bibitem [{\citenamefont {Xu}\ \emph {et~al.}(2014)\citenamefont {Xu},
  \citenamefont {Yao}, \citenamefont {Xiao},\ and\ \citenamefont
  {Heinz}}]{Xu2014}%
  \BibitemOpen
  \bibfield  {author} {\bibinfo {author} {\bibfnamefont {X.}~\bibnamefont
  {Xu}}, \bibinfo {author} {\bibfnamefont {W.}~\bibnamefont {Yao}}, \bibinfo
  {author} {\bibfnamefont {D.}~\bibnamefont {Xiao}}, \ and\ \bibinfo {author}
  {\bibfnamefont {T.~F.}\ \bibnamefont {Heinz}},\ }\href
  {https://dx.doi.org/10.1038/nphys2942} {\bibfield  {journal} {\bibinfo
  {journal} {Nat Phys}\ }\textbf {\bibinfo {volume} {10}},\ \bibinfo {pages}
  {343} (\bibinfo {year} {2014})},\ \bibinfo {note} {review}\BibitemShut
  {NoStop}%
\bibitem [{\citenamefont {Das}\ and\ \citenamefont
  {Dolui}(2015)}]{PhysRevB.91.094510}%
  \BibitemOpen
  \bibfield  {author} {\bibinfo {author} {\bibfnamefont {T.}~\bibnamefont
  {Das}}\ and\ \bibinfo {author} {\bibfnamefont {K.}~\bibnamefont {Dolui}},\
  }\href {\doibase 10.1103/PhysRevB.91.094510} {\bibfield  {journal} {\bibinfo
  {journal} {Phys. Rev. B}\ }\textbf {\bibinfo {volume} {91}},\ \bibinfo
  {pages} {094510} (\bibinfo {year} {2015})}\BibitemShut {NoStop}%
\bibitem [{\citenamefont {Lee}\ \emph {et~al.}(2015)\citenamefont {Lee},
  \citenamefont {Mak},\ and\ \citenamefont {Shan}}]{1508.03068}%
  \BibitemOpen
  \bibfield  {author} {\bibinfo {author} {\bibfnamefont {J.}~\bibnamefont
  {Lee}}, \bibinfo {author} {\bibfnamefont {K.~F.}\ \bibnamefont {Mak}}, \ and\
  \bibinfo {author} {\bibfnamefont {J.}~\bibnamefont {Shan}},\ }\href
  {https://arxiv.org/abs/1508.03068} {\  (\bibinfo {year} {2015})},\ \Eprint
  {http://arxiv.org/abs/1508.03068} {arXiv:1508.03068 [cond-mat.mes-hall]}
  \BibitemShut {NoStop}%
\bibitem [{\citenamefont {Bromley}\ \emph {et~al.}(1972)\citenamefont
  {Bromley}, \citenamefont {Murray},\ and\ \citenamefont
  {Yoffe}}]{0022-3719-5-7-007}%
  \BibitemOpen
  \bibfield  {author} {\bibinfo {author} {\bibfnamefont {R.~A.}\ \bibnamefont
  {Bromley}}, \bibinfo {author} {\bibfnamefont {R.~B.}\ \bibnamefont {Murray}},
  \ and\ \bibinfo {author} {\bibfnamefont {A.~D.}\ \bibnamefont {Yoffe}},\
  }\href {http://stacks.iop.org/0022-3719/5/i=7/a=007} {\bibfield  {journal}
  {\bibinfo  {journal} {Journal of Physics C: Solid State Physics}\ }\textbf
  {\bibinfo {volume} {5}},\ \bibinfo {pages} {759} (\bibinfo {year}
  {1972})}\BibitemShut {NoStop}%
\bibitem [{\citenamefont {B\"oker}\ \emph {et~al.}(2001)\citenamefont
  {B\"oker}, \citenamefont {Severin}, \citenamefont {M\"uller}, \citenamefont
  {Janowitz}, \citenamefont {Manzke}, \citenamefont {Vo\ss{}}, \citenamefont
  {Kr\"uger}, \citenamefont {Mazur},\ and\ \citenamefont
  {Pollmann}}]{PhysRevB.64.235305}%
  \BibitemOpen
  \bibfield  {author} {\bibinfo {author} {\bibfnamefont {T.}~\bibnamefont
  {B\"oker}}, \bibinfo {author} {\bibfnamefont {R.}~\bibnamefont {Severin}},
  \bibinfo {author} {\bibfnamefont {A.}~\bibnamefont {M\"uller}}, \bibinfo
  {author} {\bibfnamefont {C.}~\bibnamefont {Janowitz}}, \bibinfo {author}
  {\bibfnamefont {R.}~\bibnamefont {Manzke}}, \bibinfo {author} {\bibfnamefont
  {D.}~\bibnamefont {Vo\ss{}}}, \bibinfo {author} {\bibfnamefont
  {P.}~\bibnamefont {Kr\"uger}}, \bibinfo {author} {\bibfnamefont
  {A.}~\bibnamefont {Mazur}}, \ and\ \bibinfo {author} {\bibfnamefont
  {J.}~\bibnamefont {Pollmann}},\ }\href {\doibase 10.1103/PhysRevB.64.235305}
  {\bibfield  {journal} {\bibinfo  {journal} {Phys. Rev. B}\ }\textbf {\bibinfo
  {volume} {64}},\ \bibinfo {pages} {235305} (\bibinfo {year}
  {2001})}\BibitemShut {NoStop}%
\bibitem [{\citenamefont {Mak}\ \emph {et~al.}(2010)\citenamefont {Mak},
  \citenamefont {Lee}, \citenamefont {Hone}, \citenamefont {Shan},\ and\
  \citenamefont {Heinz}}]{PhysRevLett.105.136805}%
  \BibitemOpen
  \bibfield  {author} {\bibinfo {author} {\bibfnamefont {K.~F.}\ \bibnamefont
  {Mak}}, \bibinfo {author} {\bibfnamefont {C.}~\bibnamefont {Lee}}, \bibinfo
  {author} {\bibfnamefont {J.}~\bibnamefont {Hone}}, \bibinfo {author}
  {\bibfnamefont {J.}~\bibnamefont {Shan}}, \ and\ \bibinfo {author}
  {\bibfnamefont {T.~F.}\ \bibnamefont {Heinz}},\ }\href {\doibase
  10.1103/PhysRevLett.105.136805} {\bibfield  {journal} {\bibinfo  {journal}
  {Phys. Rev. Lett.}\ }\textbf {\bibinfo {volume} {105}},\ \bibinfo {pages}
  {136805} (\bibinfo {year} {2010})}\BibitemShut {NoStop}%
\bibitem [{\citenamefont {Splendiani}\ \emph {et~al.}(2010)\citenamefont
  {Splendiani}, \citenamefont {Sun}, \citenamefont {Zhang}, \citenamefont {Li},
  \citenamefont {Kim}, \citenamefont {Chim}, \citenamefont {Galli},\ and\
  \citenamefont {Wang}}]{doi:10.1021/nl903868w}%
  \BibitemOpen
  \bibfield  {author} {\bibinfo {author} {\bibfnamefont {A.}~\bibnamefont
  {Splendiani}}, \bibinfo {author} {\bibfnamefont {L.}~\bibnamefont {Sun}},
  \bibinfo {author} {\bibfnamefont {Y.}~\bibnamefont {Zhang}}, \bibinfo
  {author} {\bibfnamefont {T.}~\bibnamefont {Li}}, \bibinfo {author}
  {\bibfnamefont {J.}~\bibnamefont {Kim}}, \bibinfo {author} {\bibfnamefont
  {C.-Y.}\ \bibnamefont {Chim}}, \bibinfo {author} {\bibfnamefont
  {G.}~\bibnamefont {Galli}}, \ and\ \bibinfo {author} {\bibfnamefont
  {F.}~\bibnamefont {Wang}},\ }\href {\doibase 10.1021/nl903868w} {\bibfield
  {journal} {\bibinfo  {journal} {Nano Letters}\ }\textbf {\bibinfo {volume}
  {10}},\ \bibinfo {pages} {1271} (\bibinfo {year} {2010})},\ \bibinfo {note}
  {pMID: 20229981}\BibitemShut {NoStop}%
\bibitem [{\citenamefont {Korm\'anyos}\ \emph {et~al.}(2013)\citenamefont
  {Korm\'anyos}, \citenamefont {Z\'olyomi}, \citenamefont {Drummond},
  \citenamefont {Rakyta}, \citenamefont {Burkard},\ and\ \citenamefont
  {Fal'ko}}]{PhysRevB.88.045416}%
  \BibitemOpen
  \bibfield  {author} {\bibinfo {author} {\bibfnamefont {A.}~\bibnamefont
  {Korm\'anyos}}, \bibinfo {author} {\bibfnamefont {V.}~\bibnamefont
  {Z\'olyomi}}, \bibinfo {author} {\bibfnamefont {N.~D.}\ \bibnamefont
  {Drummond}}, \bibinfo {author} {\bibfnamefont {P.}~\bibnamefont {Rakyta}},
  \bibinfo {author} {\bibfnamefont {G.}~\bibnamefont {Burkard}}, \ and\
  \bibinfo {author} {\bibfnamefont {V.~I.}\ \bibnamefont {Fal'ko}},\ }\href
  {\doibase 10.1103/PhysRevB.88.045416} {\bibfield  {journal} {\bibinfo
  {journal} {Phys. Rev. B}\ }\textbf {\bibinfo {volume} {88}},\ \bibinfo
  {pages} {045416} (\bibinfo {year} {2013})}\BibitemShut {NoStop}%
\bibitem [{\citenamefont {Liu}\ \emph {et~al.}(2013)\citenamefont {Liu},
  \citenamefont {Shan}, \citenamefont {Yao}, \citenamefont {Yao},\ and\
  \citenamefont {Xiao}}]{PhysRevB.88.085433}%
  \BibitemOpen
  \bibfield  {author} {\bibinfo {author} {\bibfnamefont {G.-B.}\ \bibnamefont
  {Liu}}, \bibinfo {author} {\bibfnamefont {W.-Y.}\ \bibnamefont {Shan}},
  \bibinfo {author} {\bibfnamefont {Y.}~\bibnamefont {Yao}}, \bibinfo {author}
  {\bibfnamefont {W.}~\bibnamefont {Yao}}, \ and\ \bibinfo {author}
  {\bibfnamefont {D.}~\bibnamefont {Xiao}},\ }\href {\doibase
  10.1103/PhysRevB.88.085433} {\bibfield  {journal} {\bibinfo  {journal} {Phys.
  Rev. B}\ }\textbf {\bibinfo {volume} {88}},\ \bibinfo {pages} {085433}
  (\bibinfo {year} {2013})}\BibitemShut {NoStop}%
\bibitem [{\citenamefont {Xiao}\ \emph {et~al.}(2010)\citenamefont {Xiao},
  \citenamefont {Chang},\ and\ \citenamefont {Niu}}]{RevModPhys.82.1959}%
  \BibitemOpen
  \bibfield  {author} {\bibinfo {author} {\bibfnamefont {D.}~\bibnamefont
  {Xiao}}, \bibinfo {author} {\bibfnamefont {M.-C.}\ \bibnamefont {Chang}}, \
  and\ \bibinfo {author} {\bibfnamefont {Q.}~\bibnamefont {Niu}},\ }\href
  {\doibase 10.1103/RevModPhys.82.1959} {\bibfield  {journal} {\bibinfo
  {journal} {Rev. Mod. Phys.}\ }\textbf {\bibinfo {volume} {82}},\ \bibinfo
  {pages} {1959} (\bibinfo {year} {2010})}\BibitemShut {NoStop}%
\bibitem [{\citenamefont {Xiao}\ \emph {et~al.}(2012)\citenamefont {Xiao},
  \citenamefont {Liu}, \citenamefont {Feng}, \citenamefont {Xu},\ and\
  \citenamefont {Yao}}]{PhysRevLett.108.196802}%
  \BibitemOpen
  \bibfield  {author} {\bibinfo {author} {\bibfnamefont {D.}~\bibnamefont
  {Xiao}}, \bibinfo {author} {\bibfnamefont {G.-B.}\ \bibnamefont {Liu}},
  \bibinfo {author} {\bibfnamefont {W.}~\bibnamefont {Feng}}, \bibinfo {author}
  {\bibfnamefont {X.}~\bibnamefont {Xu}}, \ and\ \bibinfo {author}
  {\bibfnamefont {W.}~\bibnamefont {Yao}},\ }\href {\doibase
  10.1103/PhysRevLett.108.196802} {\bibfield  {journal} {\bibinfo  {journal}
  {Phys. Rev. Lett.}\ }\textbf {\bibinfo {volume} {108}},\ \bibinfo {pages}
  {196802} (\bibinfo {year} {2012})}\BibitemShut {NoStop}%
\bibitem [{\citenamefont {Mak}\ \emph {et~al.}(2014)\citenamefont {Mak},
  \citenamefont {McGill}, \citenamefont {Park},\ and\ \citenamefont
  {McEuen}}]{Mak27062014}%
  \BibitemOpen
  \bibfield  {author} {\bibinfo {author} {\bibfnamefont {K.~F.}\ \bibnamefont
  {Mak}}, \bibinfo {author} {\bibfnamefont {K.~L.}\ \bibnamefont {McGill}},
  \bibinfo {author} {\bibfnamefont {J.}~\bibnamefont {Park}}, \ and\ \bibinfo
  {author} {\bibfnamefont {P.~L.}\ \bibnamefont {McEuen}},\ }\href {\doibase
  10.1126/science.1250140} {\bibfield  {journal} {\bibinfo  {journal}
  {Science}\ }\textbf {\bibinfo {volume} {344}},\ \bibinfo {pages} {1489}
  (\bibinfo {year} {2014})}\BibitemShut {NoStop}%
\bibitem [{\citenamefont {Lu}\ \emph {et~al.}(2015)\citenamefont {Lu},
  \citenamefont {Zheliuk}, \citenamefont {Leermakers}, \citenamefont {Yuan},
  \citenamefont {Zeitler}, \citenamefont {Law},\ and\ \citenamefont
  {Ye}}]{Lu1353}%
  \BibitemOpen
  \bibfield  {author} {\bibinfo {author} {\bibfnamefont {J.~M.}\ \bibnamefont
  {Lu}}, \bibinfo {author} {\bibfnamefont {O.}~\bibnamefont {Zheliuk}},
  \bibinfo {author} {\bibfnamefont {I.}~\bibnamefont {Leermakers}}, \bibinfo
  {author} {\bibfnamefont {N.~F.~Q.}\ \bibnamefont {Yuan}}, \bibinfo {author}
  {\bibfnamefont {U.}~\bibnamefont {Zeitler}}, \bibinfo {author} {\bibfnamefont
  {K.~T.}\ \bibnamefont {Law}}, \ and\ \bibinfo {author} {\bibfnamefont
  {J.~T.}\ \bibnamefont {Ye}},\ }\href {\doibase 10.1126/science.aab2277}
  {\bibfield  {journal} {\bibinfo  {journal} {Science}\ }\textbf {\bibinfo
  {volume} {350}},\ \bibinfo {pages} {1353} (\bibinfo {year}
  {2015})}\BibitemShut {NoStop}%
\bibitem [{\citenamefont {Xi}\ \emph {et~al.}(2016)\citenamefont {Xi},
  \citenamefont {Wang}, \citenamefont {Zhao}, \citenamefont {Park},
  \citenamefont {Law}, \citenamefont {Berger}, \citenamefont {Forro},
  \citenamefont {Shan},\ and\ \citenamefont {Mak}}]{Xi2016}%
  \BibitemOpen
  \bibfield  {author} {\bibinfo {author} {\bibfnamefont {X.}~\bibnamefont
  {Xi}}, \bibinfo {author} {\bibfnamefont {Z.}~\bibnamefont {Wang}}, \bibinfo
  {author} {\bibfnamefont {W.}~\bibnamefont {Zhao}}, \bibinfo {author}
  {\bibfnamefont {J.-H.}\ \bibnamefont {Park}}, \bibinfo {author}
  {\bibfnamefont {K.~T.}\ \bibnamefont {Law}}, \bibinfo {author} {\bibfnamefont
  {H.}~\bibnamefont {Berger}}, \bibinfo {author} {\bibfnamefont
  {L.}~\bibnamefont {Forro}}, \bibinfo {author} {\bibfnamefont
  {J.}~\bibnamefont {Shan}}, \ and\ \bibinfo {author} {\bibfnamefont {K.~F.}\
  \bibnamefont {Mak}},\ }\href {https://dx.doi.org/10.1038/nphys3538}
  {\bibfield  {journal} {\bibinfo  {journal} {Nat Phys}\ }\textbf {\bibinfo
  {volume} {12}},\ \bibinfo {pages} {139} (\bibinfo {year} {2016})},\ \bibinfo
  {note} {letter}\BibitemShut {NoStop}%
\bibitem [{\citenamefont {Saito}\ \emph {et~al.}(2016)\citenamefont {Saito},
  \citenamefont {Nakamura}, \citenamefont {Bahramy}, \citenamefont {Kohama},
  \citenamefont {Ye}, \citenamefont {Kasahara}, \citenamefont {Nakagawa},
  \citenamefont {Onga}, \citenamefont {Tokunaga}, \citenamefont {Nojima},
  \citenamefont {Yanase},\ and\ \citenamefont {Iwasa}}]{Saito2016}%
  \BibitemOpen
  \bibfield  {author} {\bibinfo {author} {\bibfnamefont {Y.}~\bibnamefont
  {Saito}}, \bibinfo {author} {\bibfnamefont {Y.}~\bibnamefont {Nakamura}},
  \bibinfo {author} {\bibfnamefont {M.~S.}\ \bibnamefont {Bahramy}}, \bibinfo
  {author} {\bibfnamefont {Y.}~\bibnamefont {Kohama}}, \bibinfo {author}
  {\bibfnamefont {J.}~\bibnamefont {Ye}}, \bibinfo {author} {\bibfnamefont
  {Y.}~\bibnamefont {Kasahara}}, \bibinfo {author} {\bibfnamefont
  {Y.}~\bibnamefont {Nakagawa}}, \bibinfo {author} {\bibfnamefont
  {M.}~\bibnamefont {Onga}}, \bibinfo {author} {\bibfnamefont {M.}~\bibnamefont
  {Tokunaga}}, \bibinfo {author} {\bibfnamefont {T.}~\bibnamefont {Nojima}},
  \bibinfo {author} {\bibfnamefont {Y.}~\bibnamefont {Yanase}}, \ and\ \bibinfo
  {author} {\bibfnamefont {Y.}~\bibnamefont {Iwasa}},\ }\href
  {https://dx.doi.org/10.1038/nphys3580} {\bibfield  {journal} {\bibinfo
  {journal} {Nat Phys}\ }\textbf {\bibinfo {volume} {12}},\ \bibinfo {pages}
  {144} (\bibinfo {year} {2016})},\ \bibinfo {note} {letter}\BibitemShut
  {NoStop}%
\bibitem [{\citenamefont {Zhou}\ \emph {et~al.}(2016)\citenamefont {Zhou},
  \citenamefont {Yuan}, \citenamefont {Jiang},\ and\ \citenamefont
  {Law}}]{PhysRevB.93.180501}%
  \BibitemOpen
  \bibfield  {author} {\bibinfo {author} {\bibfnamefont {B.~T.}\ \bibnamefont
  {Zhou}}, \bibinfo {author} {\bibfnamefont {N.~F.~Q.}\ \bibnamefont {Yuan}},
  \bibinfo {author} {\bibfnamefont {H.-L.}\ \bibnamefont {Jiang}}, \ and\
  \bibinfo {author} {\bibfnamefont {K.~T.}\ \bibnamefont {Law}},\ }\href
  {\doibase 10.1103/PhysRevB.93.180501} {\bibfield  {journal} {\bibinfo
  {journal} {Phys. Rev. B}\ }\textbf {\bibinfo {volume} {93}},\ \bibinfo
  {pages} {180501} (\bibinfo {year} {2016})}\BibitemShut {NoStop}%
\bibitem [{\citenamefont {Yuan}\ \emph {et~al.}(2014)\citenamefont {Yuan},
  \citenamefont {Mak},\ and\ \citenamefont {Law}}]{PhysRevLett.113.097001}%
  \BibitemOpen
  \bibfield  {author} {\bibinfo {author} {\bibfnamefont {N.~F.~Q.}\
  \bibnamefont {Yuan}}, \bibinfo {author} {\bibfnamefont {K.~F.}\ \bibnamefont
  {Mak}}, \ and\ \bibinfo {author} {\bibfnamefont {K.~T.}\ \bibnamefont
  {Law}},\ }\href {\doibase 10.1103/PhysRevLett.113.097001} {\bibfield
  {journal} {\bibinfo  {journal} {Phys. Rev. Lett.}\ }\textbf {\bibinfo
  {volume} {113}},\ \bibinfo {pages} {097001} (\bibinfo {year}
  {2014})}\BibitemShut {NoStop}%
\bibitem [{\citenamefont {Zhang}\ and\ \citenamefont
  {Aji}(2016)}]{PhysRevB.94.060501}%
  \BibitemOpen
  \bibfield  {author} {\bibinfo {author} {\bibfnamefont {J.}~\bibnamefont
  {Zhang}}\ and\ \bibinfo {author} {\bibfnamefont {V.}~\bibnamefont {Aji}},\
  }\href {\doibase 10.1103/PhysRevB.94.060501} {\bibfield  {journal} {\bibinfo
  {journal} {Phys. Rev. B}\ }\textbf {\bibinfo {volume} {94}},\ \bibinfo
  {pages} {060501} (\bibinfo {year} {2016})}\BibitemShut {NoStop}%
\bibitem [{\citenamefont {Gor'kov}\ and\ \citenamefont
  {Rashba}(2001)}]{PhysRevLett.87.037004}%
  \BibitemOpen
  \bibfield  {author} {\bibinfo {author} {\bibfnamefont {L.~P.}\ \bibnamefont
  {Gor'kov}}\ and\ \bibinfo {author} {\bibfnamefont {E.~I.}\ \bibnamefont
  {Rashba}},\ }\href {\doibase 10.1103/PhysRevLett.87.037004} {\bibfield
  {journal} {\bibinfo  {journal} {Phys. Rev. Lett.}\ }\textbf {\bibinfo
  {volume} {87}},\ \bibinfo {pages} {037004} (\bibinfo {year}
  {2001})}\BibitemShut {NoStop}%
\bibitem [{Note1()}]{Note1}%
  \BibitemOpen
  \bibinfo {note} {Note that all sums over $\protect \mathbf {k}$ are
  restricted to $\left | \protect \mathbf {k}\right |$ less than some cutoff
  that restricts the momentum to a single valley.}\BibitemShut {Stop}%
\bibitem [{Note2()}]{Note2}%
  \BibitemOpen
  \bibinfo {note} {For odd parity interactions, where $v \left (-\protect
  \mathbf {k}\right )= -v \left (\protect \mathbf {k}\right )$, the intravalley
  pairing is not excluded by symmetry. Specifically, repeating the calculation
  with this assumption, the intervalley terms fully cancel, and one obtains
  \protect \cref {eq:channels} without the intervalley term on the third
  line.}\BibitemShut {Stop}%
\bibitem [{Note3()}]{Note3}%
  \BibitemOpen
  \bibinfo {note} {For example, using the values for \ce {WSe2}, $\protect
  \qopname \relax o{sin}^2 {{\theta }_{\protect \mathbf {k}}} = 0.44$ and
  $\protect \qopname \relax o{cos}^2 {{\theta }_{\protect \mathbf {k}}} = 0.56$
  at the chemical potential.}\BibitemShut {Stop}%
\bibitem [{Note4()}]{Note4}%
  \BibitemOpen
  \bibinfo {note} {Note that the full ground state also contains the two lower
  filled bands, but those contribute zero net Berry curvature and may be
  ignored in this section and the next.}\BibitemShut {Stop}%
\bibitem [{\citenamefont {Maki}\ and\ \citenamefont
  {Tsuneto}(1964)}]{Maki01061964}%
  \BibitemOpen
  \bibfield  {author} {\bibinfo {author} {\bibfnamefont {K.}~\bibnamefont
  {Maki}}\ and\ \bibinfo {author} {\bibfnamefont {T.}~\bibnamefont {Tsuneto}},\
  }\href {\doibase 10.1143/PTP.31.945} {\bibfield  {journal} {\bibinfo
  {journal} {Progress of Theoretical Physics}\ }\textbf {\bibinfo {volume}
  {31}},\ \bibinfo {pages} {945} (\bibinfo {year} {1964})}\BibitemShut
  {NoStop}%
\bibitem [{\citenamefont {Abrikosov}\ and\ \citenamefont
  {Gor'kov}(1961)}]{AbrikosovGorkov1961}%
  \BibitemOpen
  \bibfield  {author} {\bibinfo {author} {\bibfnamefont {A.}~\bibnamefont
  {Abrikosov}}\ and\ \bibinfo {author} {\bibfnamefont {L.}~\bibnamefont
  {Gor'kov}},\ }\href@noop {} {\bibfield  {journal} {\bibinfo  {journal}
  {Soviet Phys. JETP}\ }\textbf {\bibinfo {volume} {12}},\ \bibinfo {pages}
  {1243} (\bibinfo {year} {1961})}\BibitemShut {NoStop}%
\bibitem [{\citenamefont {Maki}(1969)}]{maki1969superconductivity}%
  \BibitemOpen
  \bibfield  {author} {\bibinfo {author} {\bibfnamefont {K.}~\bibnamefont
  {Maki}},\ }in\ \href@noop {} {\emph {\bibinfo {booktitle}
  {Superconductivity}}},\ \bibinfo {editor} {edited by\ \bibinfo {editor}
  {\bibfnamefont {R.~D.}\ \bibnamefont {Parks}}}\ (\bibinfo  {publisher}
  {Dekker, New York},\ \bibinfo {year} {1969})\BibitemShut {NoStop}%
\bibitem [{\citenamefont {Rold\'an}\ \emph {et~al.}(2013)\citenamefont
  {Rold\'an}, \citenamefont {Cappelluti},\ and\ \citenamefont
  {Guinea}}]{PhysRevB.88.054515}%
  \BibitemOpen
  \bibfield  {author} {\bibinfo {author} {\bibfnamefont {R.}~\bibnamefont
  {Rold\'an}}, \bibinfo {author} {\bibfnamefont {E.}~\bibnamefont
  {Cappelluti}}, \ and\ \bibinfo {author} {\bibfnamefont {F.}~\bibnamefont
  {Guinea}},\ }\href {\doibase 10.1103/PhysRevB.88.054515} {\bibfield
  {journal} {\bibinfo  {journal} {Phys. Rev. B}\ }\textbf {\bibinfo {volume}
  {88}},\ \bibinfo {pages} {054515} (\bibinfo {year} {2013})}\BibitemShut
  {NoStop}%
\bibitem [{Note5()}]{Note5}%
  \BibitemOpen
  \bibinfo {note} {Related software and source code at \\ \protect \url
  {https://evansosenko.com/dichalcogenides}}\BibitemShut {NoStop}%
\bibitem [{\citenamefont {Hunter}(2007)}]{Hunter:2007}%
  \BibitemOpen
  \bibfield  {author} {\bibinfo {author} {\bibfnamefont {J.~D.}\ \bibnamefont
  {Hunter}},\ }\href {\doibase 10.1109/MCSE.2007.55} {\bibfield  {journal}
  {\bibinfo  {journal} {Computing In Science \& Engineering}\ }\textbf
  {\bibinfo {volume} {9}},\ \bibinfo {pages} {90} (\bibinfo {year}
  {2007})}\BibitemShut {NoStop}%
\end{thebibliography}
\end{document}